\documentclass[letterpaper,12pt]{article}

\usepackage{amsmath}
\usepackage{amssymb}
\usepackage{booktabs}
\usepackage{graphicx}
\usepackage{gensymb}
\usepackage{subcaption}
\usepackage{lipsum}
\usepackage{cases}
\usepackage{mathtools, cuted}
\usepackage{url}

\makeatletter
\def\Arg{\mathop{\operator@font Arg}\nolimits}
\makeatother

\graphicspath{ {./} }
\usepackage{subcaption}
\usepackage[onehalfspacing]{setspace}
\usepackage{tocvsec2}
\usepackage[bookmarksdepth=subsection]{hyperref}
\usepackage{bookmark}
\usepackage[labelfont=bf]{caption}
\usepackage[justification=justified]{caption}
\usepackage[margin=1in]{geometry}

\begin{document}
\begin{titlepage}
	
	\clearpage\thispagestyle{empty}
	
	%\noindent {\footnotesize {{\em
	
	%\hfill To be submitted to Cement and Concrete Composites} }} \\
	
	\noindent
	
	\hrulefill
	
	\begin{figure}[h!]
		
		\centering
		
		\includegraphics[width=1.5 in]{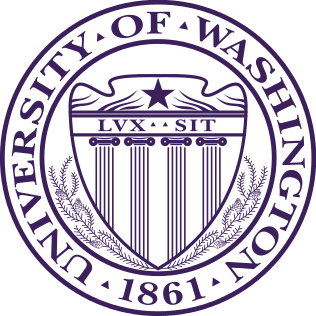}
		
	\end{figure}
	
	%{\color{NU} some text}
	
	\begin{center}
		
		{
			
			{\bf A\&A Program in Structures} \\ [0.1in]
			
			William E. Boeing Department of Aeronautics and Astronautics \\ [0.1in]
			
			University of Washington \\ [0.1in]
			
			Seattle, Washington 98195, USA

		}
		
	\end{center} %\vskip 5mm
	
	\hrulefill \\ \vskip 2mm
	
	\vskip 0.5in
	
	\begin{center}
		
		{\large {\bf Enhancing the Electrical and Thermal Conductivities of Polymer Composites via Curvilinear Fibers: An Analytical Study}}\\[0.5in]
		
		{\large {\sc Marco Salviato, Sean E. Phenisee}}\\[0.75in]

		{\sf \bf INTERNAL REPORT No. 19-02/01E}\\[0.75in]

	\end{center}

	\noindent {\footnotesize {{\em Submitted to Mathematics and Mechanics of Solids \hfill February 2019} }}
	
\end{titlepage}

\newpage
\tableofcontents
% ***** BEGIN TEXT!!***

\newpage
\begin{abstract}
	The new generation of manufacturing technologies such as e.g. additive manufacturing and automated fiber placement has enabled the development of material systems with desired functional and mechanical properties via particular designs of inhomogeneities and their mesostructural arrangement. Among these systems, particularly interesting are materials exhibiting \textbf{Curvilinear Transverse Isotropy} (CTI) in which the inhomogeneities take the form of continuous fibers following curvilinear paths designed to e.g. optimize the electric and thermal conductivity, and the mechanical performance of the system. In this context, the present work proposes a general framework for the exact, closed-form solution of electrostatic problems in materials featuring curvilinear transverse isotropy. First, the general equations for the fiber paths that optimize the	electric conductivity are derived leveraging a proper conformal coordinate system. Then, the continuity equation for the curvilinear, transversely isotropic system is derived in terms of electrostatic potential. A general exact, closed-form expression for the electrostatic potential and electric field is derived and validated by Finite Element Analysis. Finally, potential avenues for the development of materials with superior electric conductivity and damage sensing capabilities are discussed. 
\end{abstract}

\providecommand{\keywords}
{	
	\textit{\textbf{Keywords:}} 
}


\section{Introduction}

Emerging manufacturing technologies have provided engineers with an unprecedentedly broad design space. Laser lithography has enabled the development of lattice materials with sub-micron resolution \cite{Fleck10, Greer11, Greer13, Bauer14, Bauer16}[e.g.]. Multi-material additive manufacturing has allowed the design of multi-phase systems with novel physical and mechanical properties \cite{Zava15a, Zava15b, Mar17}[e.g.]. Automated fiber deposition along curvilinear paths has introduced new ways to achieve damage tolerance of composite structures \cite{Pasini11, Dirk12, Pasini14}. 

A common trait of all these technologies is that they have enabled the development of material systems with desired functional and mechanical properties via particular designs of inhomogeneities and their micro- and meso-structural arrangements \cite{wang17}. Among these systems, particularly interesting are materials exhibiting \textit{Curvilinear Transverse Isotropy} (CTI) in which the inhomogeneities take the form of continuous fibers following curvilinear paths. The morphology of the fiber curves is designed to e.g. optimize the electric and thermal conductivity, and the mechanical performance of the system \cite{Nik15,Dod16}.

\begin{figure*}
	\center
	\includegraphics[trim=0cm 0cm 0cm 0cm, clip=true,clip=true,width = 0.8\textwidth]{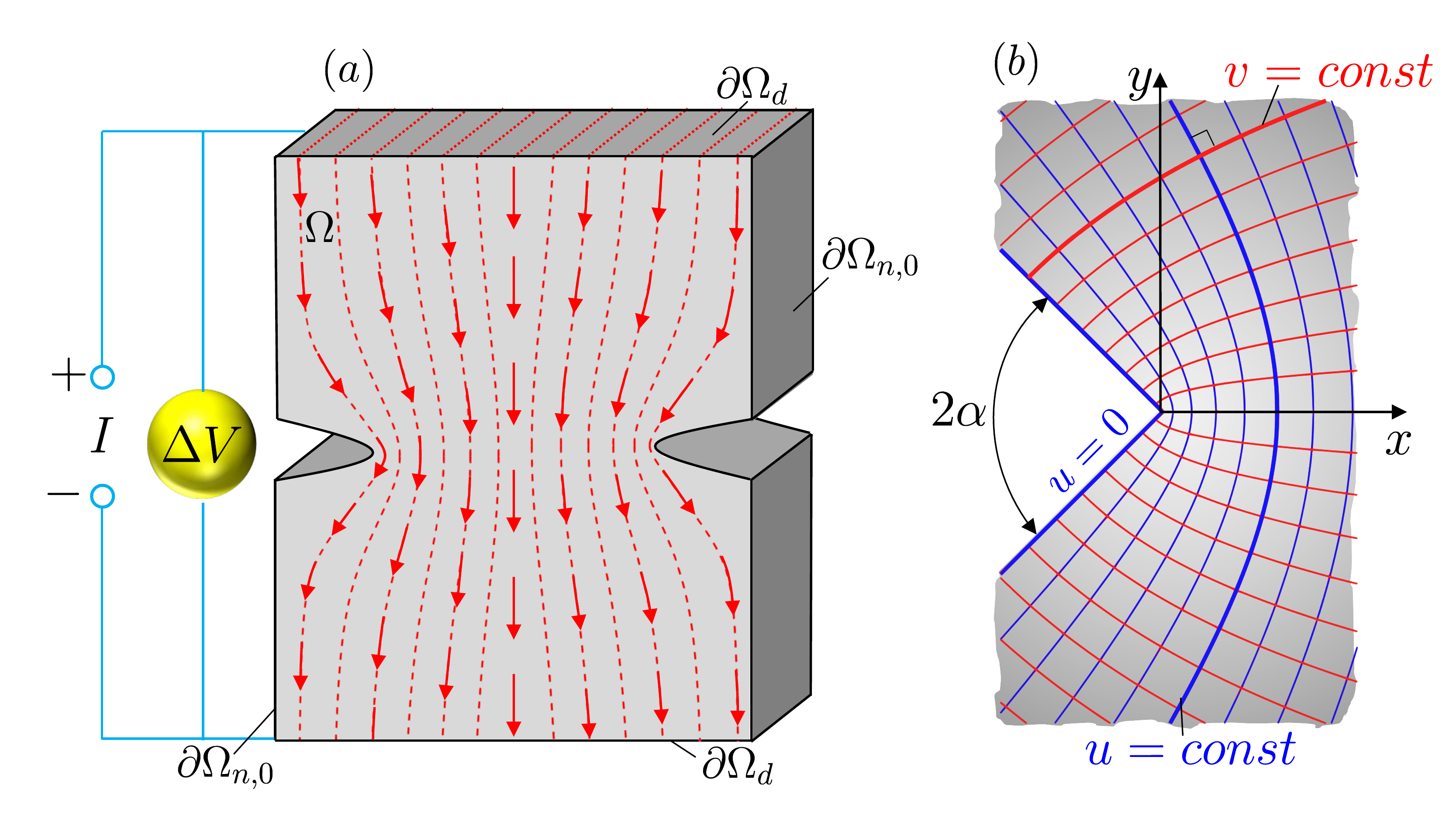}
	\caption{(a) a notched $2$D body $\Omega\cup \partial \Omega$ subjected to an electric potential and (b) example of a conformal mapping used in this work (equation (\ref{eq2.39})).}
	\label{f1}
\end{figure*}

In this context, the present work investigates the use of curvilinear conductive fibers to increase the conductivity of insulating materials such as e.g. polymer or concrete. Achieving a sufficient electrical conductivity is quintessential to enable the use of resistivity measurements for the detection of damage initiation and evolution close to stress raisers such as sharp notches, corners and cracks. Towards this goal, several researchers investigated the use of e.g. nanoparticles or graphene to increase the conductivity \cite{Le14,Pie12,DuXuLiu16}[e.g.] but the application of curvilinear fibers has never been explored. This manuscript takes a first step in this direction by proposing a general framework for the exact, closed-form solution of electrostatic problems in materials featuring curvilinear transverse isotropy. First, the general equations for the fiber paths optimizing the electrical conductivity are derived leveraging a proper conformal coordinate system. Then, the continuity equation for the curvilinear, transversely isotropic system is derived in terms of electrostatic potential. A general exact, closed-form expression for the electrostatic potential and electric field is derived and validated by Finite Element Analysis.

%%%%%%%%%%%%%%%%%%%%%%%%%%%%%%%%%%%%%%%%%%%%%%%%%%%%%%%%%%%%%%%%%%%%%%
\section{Preliminary remarks}\label{remarks}
%%%%%%%%%%%%%%%%%%%%%%%%%%%%%%%%%%%%%%%%%%%%%%%%%%%%%%%%%%%%%%%%%%%%%%
\subsection{Continuity equation in isotropic and homogeneous media}

Let us consider a $2$D body $\Omega\cup \partial \Omega$ made of a homogenous and isotropic material obeying the theory of the linear electrostatics and suppose that the body is subjected to a potential difference as shown in Figure \ref{f1}a. With reference to the Cartesian coordinate system ($x$,$y$,$z$) represented in Figure \ref{f1}, the equation of continuity for the electrostatic field, assuming the absence of sources, states:

\begin{equation}
\frac{\partial J_x}{\partial x}+\frac{\partial J_y}{\partial y}=-\frac{\partial \rho}{\partial t}
\label{eq2.1}
\end{equation}
where $J_i\left(i=x,y\right)=$ components of current density in $x-$ and $y-$ directions, and $\rho=$ electric charge density. The current density is related to the electric field, $E$, through Ohm's law \cite{Ohm27} $E_i=k^{-1} J_i\left(i=x,y\right)$ where $k=k_x=k_y$ represents the electrical conductivity of the material, treated as homogeneous and isotropic. Further, the conservativeness of the electric field leads to the following link to the electric potential $\varphi$: $E_x=-\partial \varphi/\partial x$ and $E_y=-\partial \varphi/\partial y$. Substituting the foregoing expressions into equation \ref{eq2.1} and focusing on the steady-state solution of the problem, one gets the following two-dimensional Laplace equation for the electric potential:
\begin{equation}
\nabla^2 \varphi=0
\label{eq2.2}
\end{equation}
with $\nabla^2=\partial^2/\partial x^2+\partial^2/\partial y^2$ being the laplacian operator. The solution of equation (\ref{eq2.2}) in a given two-dimensional domain $\Omega \cup \partial \Omega$ subjected to Dirichlet or von Neumann conditions on the boundary $\partial \Omega$, allows the full characterization of the electric potential and the related current density. Towards this goal, it is often convenient to employ a transformation of coordinates to simplify the application of the boundary conditions when the shape of the domain $\Omega$ is complex. In this work, only conformal mappings are considered, i.e. transformations that can be described by means of a complex analytic function \cite[e.g.]{Fish99,BroChu13} (an example of conformal mapping used in the present work is provided in Figure \ref{f1}b). In this case, the change of coordinates $z=z\left(\xi\right)$ with $z=x+\mbox{i}y$ and $\xi=u+\mbox{i}v$ is always bijective, satisfies the Cauchy-Riemann (C--R) conditions $\partial u/\partial x=\partial v/\partial y$, $\partial u/\partial y=-\partial v/\partial x$ in $\Omega$, and transforms the governing equation to an expression that is formally identical to the Cartesian case \cite[e.g.]{Lea15, Shep32, WigSte39a, Wig39, Gre2000, Ru00, Baz04, SalZap16, SalZap18, AnsNel18,biot56,biot59}:      
\begin{equation}
\nabla^2 \varphi=\frac{\partial^2 \varphi}{\partial u^2} + \frac{\partial^2 \varphi}{\partial v^2}= 0
\label{eq2.3}
\end{equation}
with an easier description of the boundary of the problem. In this work, the following conformal mappings are considered in which the condition $u=u_0$ or $v=v_0$ (with $u_0,v_0$ being real constants) describes the boundary $\partial \Omega_{n,0}$ \cite{SalZap16, SalZap18}. Here $\partial \Omega_{n,0}$ refers to the portion of the boundary where $J_n=\textbf{\textit{J}}\cdot \textbf{\textit{n}}=0$. 

In the introduced coordinate system, the electric field can be calculated as follows:
\begin{equation}
\textbf{\textit{E}} = -\nabla \varphi = -\frac{1}{h_u} \frac{\partial \varphi}{\partial u} \textbf{e}_u - \frac{1}{h_v} \frac{\partial \varphi}{\partial v} \textbf{e}_v 
\label{eq2.4}
\end{equation}
where $\textbf{e}_u, \textbf{e}_v$ represent the orthonormal basis of the curvilinear system. \\
$h_u=\sqrt{(\partial x/ \partial u)^2 + (\partial y/\partial u)^2}$ and $h_v=\sqrt{(\partial x/\partial v)^2 + (\partial y/ \partial v)^2}$ are the related metric coefficients.

%%%%%%%%%%%%%%%%%%%%%%%%%%%%%%%%%%%%%%%%%%%%%%%%%%%%%%%%%%%%%%%%%%%%%%
\subsection{General solution}

Let us consider the domain $\Omega\cup \partial \Omega$ and a conformal map $z=z\left(\xi\right)$ with $z=x+i y$ and $\xi=u+i v$ such that $\partial \Omega_{n,0}$ is described by the condition $u=u_0$. The constant $u_0$ is taken as a positive number, so that the domain of integration belongs to the right half plane of the $(u,v)$ space. Under this condition, $u$ is always positive whilst $v$ can take any value. The body is subjected to a uniform current density or potential in the remaining portion of the domain, $\Omega_d$ as shown in Figure \ref{f1}. 

\begin{figure*}
	\center 
	\includegraphics[trim=0cm 0cm 0cm 0cm, clip=true,clip=true,width = 0.9\textwidth]{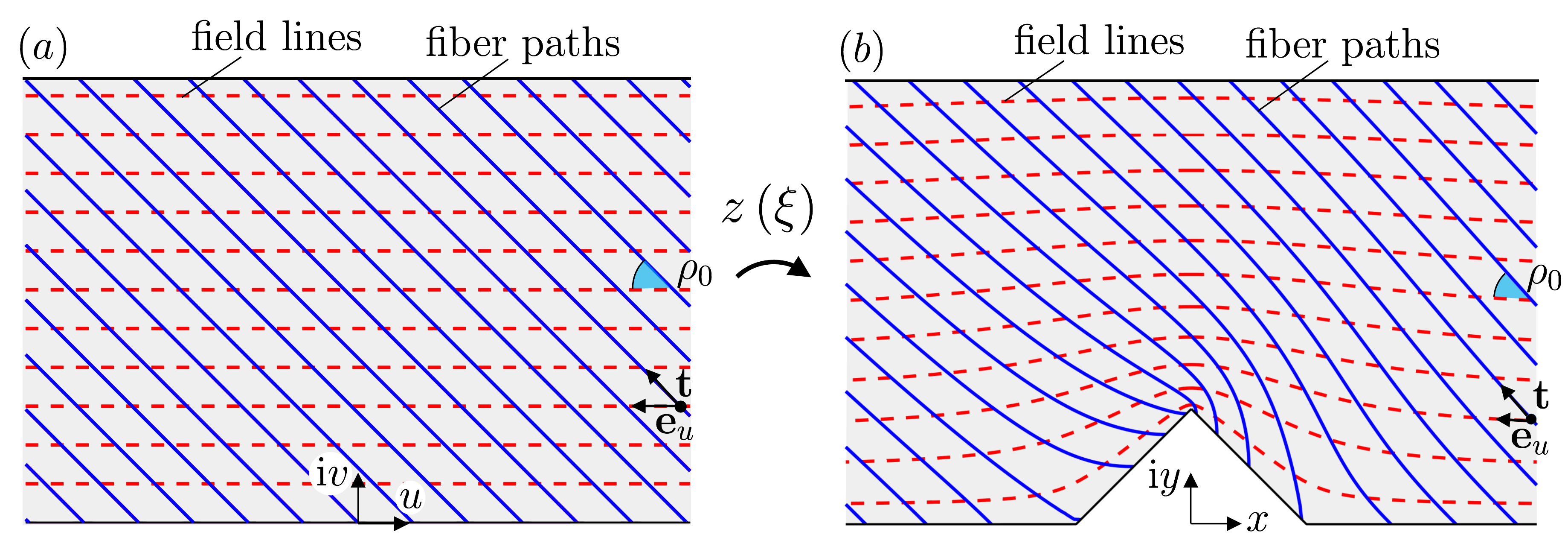}
	\caption{Schematic representation of the field lines and the family of curves $\gamma\left(s,u_c,v_c\right)$ in the (a) transformed domain $\Omega'$ and (b) Cartesian domain $\Omega$. The angle $\rho_0$ between the tangent unit vector of the field lines, in this case $\textbf{e}_u$, and of $\gamma\left(s,u_c,v_c\right)$, $\textbf{t}$, is preserved by the conformal transformation.}
	\label{f2}
\end{figure*}

Assuming separation of variables in curvilinear coordinates, $\varphi\left(u,v\right)=f\left(u\right)g\left(v\right)$, and substituting into equation (\ref{eq2.3}), one gets:
\begin{equation}
f''\left(u\right)g\left(v\right) + f\left(u\right)g''\left(v\right)=0 
\label{eq2.5}
\end{equation}
or, equivalently:
\begin{equation}
\frac{ f''\left(u\right)}{f\left(u\right)}=-\frac{g''\left(v\right)}{g\left(v\right)}=\lambda^2 
\label{eq2.6}
\end{equation}
with $\lambda$ being a real constant. Introducing the boundary conditions, one obtains the following system of differential equations:
\begin{subnumcases}{}
g''\left(v\right)+\lambda^2 g\left(v\right)=0 \label{eq2.7a} \\
f''\left(u\right)-\lambda^2 f\left(u\right)=0 \label{eq2.7b} \\
f'\left(u_0\right)=0 \label{eq2.7c} \\  
\frac{\mid f'\left(u\right)\mid}{h_u}<\infty\quad\mbox{for}\quad u,v\rightarrow \infty \label{eq2.7d} \\ 
\frac{\mid g'\left(u\right)\mid}{h_v}<\infty\quad\mbox{for}\quad u,v\rightarrow \infty \label{eq2.7e}
\end{subnumcases}
where equation (\ref{eq2.7c}) states that the component of the current density normal to the surface is zero whereas equations (\ref{eq2.7d}, \ref{eq2.7e}) express the condition for bounded, remote current. The case $\lambda^2 \neq 0$ can be disregarded since it only provides trivial solutions. In contrast, under the condition $\lambda^2 = 0$, the general solution is: 
\begin{equation}
\varphi \left(u,v\right)=\left(A+Bu\right)\left(C+Dv\right)  
\label{eq2.8}
\end{equation}
which simplifies further after applying the boundary conditions:
\begin{equation}
\varphi \left(u,v\right)=C_1 + \zeta v  
\label{eq2.9}
\end{equation}
where $C_1$ represents a reference potential and can be taken as zero whereas $\zeta$ is a real constant that can be found imposing the remote potential or current density. Then, recalling equation (\ref{eq2.4}), the curvilinear components of the current density can be written as follows:
\begin{equation}
J_u=0,\quad~~J_v=\frac{\psi}{\|z'\left(\xi\right)\|}  
\label{eq2.9bis}
\end{equation}
where $\|z'\left(\xi\right)\|=h$ is the magnitude of the first derivative of the conformal map and $\psi= -\zeta k$ is a constant to determine. Leveraging the relation between the electric field, the electric potential, and the current density along with the C--R conditions on the curvilinear coordinates one can write the Cartesian components of the current density as $J_x=\psi \partial v/\partial x$ and $J_y=\psi \partial u/\partial x$. It is worth noting that, since $\xi'=\mbox{d} \xi/\mbox{d}z = \partial u/ \partial x + i \partial v/ \partial x$, the Cartesian components of the current density represent the real and imaginary parts of the first derivative of the conformal map used to describe the domain, $\xi=\xi\left(z\right)$ \cite{SalZap16, SalZap18, ZapSal18}. Accordingly, the current density can be written as:     
\begin{equation}
J_y+i J_x=\psi \frac{\mbox{d} \xi\left(z\right)}{\mbox{d} z}  
\label{eq2.10}
\end{equation}
This result provides an explicit link between the domain geometry and the current density field. Once the proper conformal map is determined, the problem of calculating the electric potential and the related current density can be solved directly. It is worth mentioning that in case $\partial \Omega_{n,0}$ is described by the condition $v=v_0$, a similar approach can be used to show that $\varphi \left(u,v\right)=C_1 + C_2 u $. In such a case, the current density in curvilinear components can be calculated as follows:
\begin{equation}
J_u=\frac{\psi}{\|z'\left(\xi\right)\|},\quad~~J_v=0  
\label{eq2.9bisbis}
\end{equation}
whereas the Cartesian components take the following form \cite{SalZap16, SalZap18, ZapSal18}:
\begin{equation}
J_x-i J_y=\psi \frac{\mbox{d} \xi\left(z\right)}{\mbox{d} z}  
\label{eq2.11}
\end{equation}

%%%%%%%%%%%%%%%%%%%%%%%%%%%%%%%%%%%%%%%%%%%%%%%%%%%%%%%%%%%%%%%%%%%%%%
\subsection{General field line equations}
The theoretical framework described in the foregoing sections can be used to calculate the general equations of the field lines. This is a particularly important information since, as it will be shown in the following sections, the field lines also represent the fiber paths optimizing the electrical conductivity of the system. Leveraging the curvilinear, orthogonal coordinate system and considering the case in which $u=u_0$ describes $\partial \Omega_{n,0}$, this task is particularly easy as the paths are described by the condition $u=const$. In fact, being the gradient of the electric potential, $\textbf{E}$ is orthogonal to the equipotential lines which, as shown in equation (\ref{eq2.9}), are described by the condition $v=const$.

For the sake of generality, however, let us consider the family of paths defined by the condition $\textbf{t} \cdot \textbf{e}_v=\cos \rho_0$ with $\textbf{t}=$ tangent unit vector of the path, $\textbf{e}_v=$ tangent unit vector of the field lines when $u=u_0$ describes the insulated portion of the boundary and $\rho_0$ being a constant angle as represented in Figure \ref{f2}. Then, the fiber paths can be calculated as the integral curve, $\gamma\left(s\right)$, of the tangent vector $\textbf{t}=\cos \rho_0 \textbf{e}_v + \sin \rho_0 \textbf{e}_u$:
\begin{subnumcases}{}
\frac{\partial x}{\partial u} u' + \frac{\partial x}{\partial v}v' =\sin \rho_0 \frac{\partial x}{\partial u} + \cos \rho_0 \frac{\partial x}{\partial v} \label{eq2.12a} \\
\frac{\partial y}{\partial u} u' + \frac{\partial y}{\partial v}v' =\sin \rho_0 \frac{\partial y}{\partial u} + \cos \rho_0 \frac{\partial y}{\partial v} \label{eq2.12b}
\end{subnumcases}
where $u'=\mbox{d}u/\mbox{d}s$ and $v'=\mbox{d}v/\mbox{d}s$ respectively. The foregoing system was obtained noting that (a) $\textbf{e}_u=1/h_u\left(\partial x/\partial u \textbf{i}+ \partial y/\partial u \textbf{j}\right)$ and $\textbf{e}_v=1/h_v\left(\partial x/\partial v \textbf{i}+ \partial y/\partial v \textbf{j}\right)$ with $\textbf{i}$ and $\textbf{j}$ being the Cartesian unit vectors in $x-$ and $y-$direction, (b) $h_u=h_v=h$ thanks to the C--R conditions on the conformal transformation and (c) $h \textbf{e}_v$ is parallel to $\textbf{e}_v$. The solution of the foregoing system can be obtained noting that $u'=\sin \rho_0$ and $v'=\cos \rho_0$, which leads to the following solution:
\begin{equation}
\gamma\left(s,u_c,v_c\right):\quad u\left(s\right)=u_c+\sin \rho_0 s, \quad v\left(s\right)=v_c+\cos \rho_0 s 
\label{eq2.13}
\end{equation} 
where $s\in \left[0,\infty\right)$, and $u_c=u\left(0\right)$, $v_c=v\left(0\right)$ respectively. Then, substitution of equation (\ref{eq2.13}) into the conformal map, leads to the description of the curves in Cartesian coordinates: $z\left(s\right)=\left(z \circ\xi\right)\left(s\right)$. A comparison between the field lines and the family of curves $\gamma\left(s,u_c,v_c\right)$ can be found in Figure \ref{f2}.

It is worth noting here that a similar procedure can be used when $v=v_0$ describes $\partial \Omega_{n,0}$, which is common when the conformal mapping is constructed leveraging a Schwarz-Christoffel transformation \cite{DriTre02}. In such a case, the field lines are described by the condition $v=const$ and $\textbf{e}_u$ represents the unit tangent vector. The integral curves characterized by the unit vector $\textbf{t}~| ~\textbf{t}\cdot \textbf{e}_u=\cos \rho_0$ can be calculated as follows:
\begin{equation}
\gamma\left(s,u_c,v_c\right):\quad u\left(s\right)=u_c+\cos \rho_0 s, \quad v\left(s\right)=v_c+\sin \rho_0 s 
\label{eq2.14}
\end{equation}

Finally, it should be noted that when $\rho_0=0$, both equations (\ref{eq2.13}) and (\ref{eq2.14}) reduce to the equations for the field lines.

\begin{figure*}
	\center 
	\includegraphics[trim=0cm 0cm 0cm 0cm, clip=true,clip=true,width = 1\textwidth]{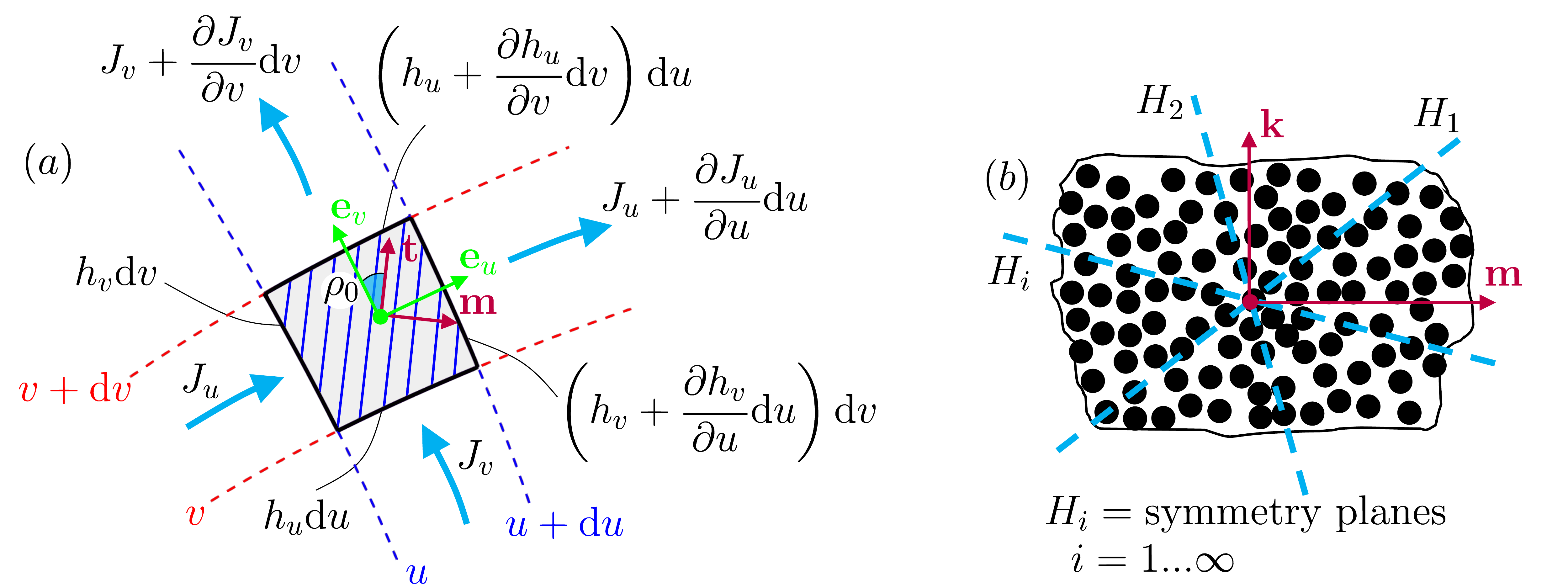}
	\caption{(a) Infinitesimal material element featuring curvilinear orthotropy in $\left(\textbf{e}_u, \textbf{e}_v,\textbf{k}\right)$ and the related current densities; (b) schematic representation of the microstructure as cut by the plane $\pi:\pi \perp \textbf{t}$. As can be noted, an infinite number of symmetry planes $H_i$ exist supporting the assumption of \textit{curvilinear transverse isotropy} in $\left(\textbf{t}, \textbf{m},\textbf{k}\right)$.}
	\label{f3}
\end{figure*}

%%%%%%%%%%%%%%%%%%%%%%%%%%%%%%%%%%%%%%%%%%%%%%%%%%%%%%%%%%%%%%%%%%%%%%
\section{Electrostatics in media featuring curvilinear transverse isotropy}\label{CTI}

%%%%%%%%%%%%%%%%%%%%%%%%%%%%%%%%%%%%%%%%%%%%%%%%%%%%%%%%%%%%%%%%%%%%%%
\subsection{Constitutive behavior}
Let us consider a material reinforced by fibers following a family of curvilinear paths $\gamma\left(s,u_c,v_c\right)$ as shown in Figure \ref{f2}. As discussed in Section \ref{remarks}, the paths are constructed such that their tangent, $\textbf{t}$ is rotated of a constant angle $\rho_0$ with respect to field lines. In the particular case in which $\rho_0=0$, the fiber paths coincide with the field lines of the problem. For each material point, a curvilinear orthonormal basis $\left(\textbf{t}, \textbf{m}, \textbf{k}\right)$ can be defined such that $\textbf{m}\times \textbf{t}= \textbf{k}$ with $\textbf{k}$ being the unit vector of the $z-$direction. Now, considering the material microstructure, it is easy to note that any plane $H: \textbf{t}\in H$ is a symmetry plane (Figure \ref{f3}b). Accordingly, in each material point the transverse plane $\pi: \pi \perp \textbf{t}$ is a plane of isotropy and the material exhibits \textit{curvilinear transverse isotropy} \cite{Sadd09,Bow09}. Assuming that each material point is reinforced by a sufficiently large number of curvilinear fibers, the material can be treated as homogeneous. Then, the anisotropic Ohm's equation can be written in matrix form as follows:
\begin{equation}
\left \{
\begin{matrix}
J_t  \\
J_m  
\end{matrix} \right\}=
\begin{bmatrix}
k_{tt} & 0 \\
0 & k_{mm} 
\end{bmatrix}\left \{
\begin{matrix}
E_t  \\
E_m 
\end{matrix}\right\}
\label{eq2.15}
\end{equation}
where $J_t,J_m,E_t,E_m$ represent the components of the current density and electric field respectively in the orthonormal basis $\left(\textbf{t}, \textbf{m}, \textbf{k}\right)$. The electric conductivities $k_{tt}$ and $k_{mm}$ in the curvilinear system can be measured experimentally or estimated by means of micromechanics \cite[e.g.]{SevKac09} and depend on the electric conductivities of the fibers and the matrix as well as the volume fraction of fibers. Assuming that the electrical conductivity of the fibers is significantly larger than the matrix and considering that fibers and matrix are in parallel coupling along $\textbf{t}$ and in series coupling along $\textbf{m}$, one can conclude that generally $k_{tt}\ge k_{mm}$.

Although the orthonormal basis $\left(\textbf{t}, \textbf{m}, \textbf{k}\right)$ guarantees the simplest description of the constitutive behavior, it is convenient to recast equation (\ref{eq2.15}) in the curvilinear coordinate system $\left(\textbf{e}_u, \textbf{e}_v, \textbf{k}\right)$ which enables the simplest application of the boundary conditions without affecting the continuity equation significantly. Considering the case in which $u=u_0$ provides the description of $\partial \Omega_{n,0}$ and recalling that $\textbf{t}=\cos \rho_0 \textbf{e}_v + \sin \rho_0 \textbf{e}_u$ and $\textbf{m}=-\sin \rho_0 \textbf{e}_v + \cos \rho_0 \textbf{e}_u$, the constitutive behavior can be written as follows:
\begin{equation}
\left \{
\begin{matrix}
J_v  \\
J_u  
\end{matrix} \right\}=
\begin{bmatrix}
k_{vv}\left(\rho_0,k_{tt},k_{mm}\right) & k_{uv}\left(\rho_0,k_{tt},k_{mm}\right) \\
k_{uv}\left(\rho_0,k_{tt},k_{mm}\right) & k_{uu}\left(\rho_0,k_{tt},k_{mm}\right)
\end{bmatrix}\left \{
\begin{matrix}
E_v  \\
E_u 
\end{matrix}\right\}
\label{eq2.16}
\end{equation}
where:
%\begin{subnumcases}{}
%k_{vv}\left(\rho_0,k_{tt},k_{mm}\right)=\cos^2\left(\rho_0\right) k_{tt} + %\sin^2\left(\rho_0\right) k_{mm}    \label{eq2.17a} \\
%k_{uu}\left(\rho_0,k_{tt},k_{mm}\right)=\sin^2\left(\rho_0\right) k_{tt} + %\cos^2\left(\rho_0\right) k_{mm} \label{eq2.17b} \\
%k_{uv}\left(\rho_0,k_{tt},k_{mm}\right)=\left(k_{tt}-k_{mm}\right)\cos\rho_0 \sin\rho_0 %\label{eq2.17c}  
%\end{subnumcases}
\begin{subnumcases}{}
k_{vv}=\cos^2\left(\rho_0\right) k_{tt} + \sin^2\left(\rho_0\right) k_{mm}    \label{eq2.17a} \\
k_{uu}=\sin^2\left(\rho_0\right) k_{tt} + \cos^2\left(\rho_0\right) k_{mm} \label{eq2.17b} \\
k_{uv}=\left(k_{tt}-k_{mm}\right)\cos\rho_0 \sin\rho_0 \label{eq2.17c}  
\end{subnumcases}
It is worth mentioning that equations (\ref{eq2.17a}), (\ref{eq2.17b}) and (\ref{eq2.17c}) can be obtained noting that $\textbf{K}^{(u,v)}=\textbf{T}^\top \textbf{K}^{(t,m)}\textbf{T}$ with $\textbf{T}$ being the rotation matrix linking $\left(\textbf{t}, \textbf{m}, \textbf{k}\right)$ to $\left(\textbf{e}_u, \textbf{e}_v, \textbf{k}\right)$ and $\textbf{K}^{(t,m)}$, $\textbf{K}^{(u,v)}$ being the conductivity matrices in the two orthonormal bases respectively.

In case that $v=v_0$ provides the description of $\partial \Omega_{n,0}$, $\textbf{e}_u$ becomes the unit tangent vector of the field lines and $\rho_0$ represents the angle between $\textbf{t}$ and $\textbf{e}_u$. It can be shown that the constitutive behavior can be obtained from equation (\ref{eq2.16}) simply by switching the diagonal terms in $\textbf{K}^{(u,v)}$.

%%%%%%%%%%%%%%%%%%%%%%%%%%%%%%%%%%%%%%%%%%%%%%%%%%%%%%%%%%%%%%%%%%%%%%
\subsection{Continuity equation}
To solve for the electric potential, it is convenient to write the continuity equation for the material system schematically represented in Figure \ref{f3} in curvilinear coordinates. Let us consider an infinitesimal curvilinear element and the current densities represented in Figure \ref{f3}. The energy balance for a infinitesimal time increment $\mbox{d}t$, excluding the presence of sources, reads:
\begin{equation} \label{eq2.18}
\begin{split}
& J_u h_u \mbox{d}v\mbox{d}t-\left(J_u+\frac{\partial J_u}{\partial u}\mbox{d}u\right)\left(h_v+\frac{\partial h_v}{\partial u}\mbox{d}u\right) \mbox{d}v\mbox{d}t  + \\
& + J_v h_v \mbox{d}u\mbox{d}t-\left(J_v+\frac{\partial J_v}{\partial v}\mbox{d}v\right)\left(h_u+\frac{\partial h_u}{\partial v}\mbox{d}v\right) \mbox{d}u\mbox{d}t=0
\end{split}
\end{equation}
which, neglecting the infinitesimal terms of higher order and recalling that $h_u=h_v=h$, $\mbox{d}t>0$, $\mbox{d}u\mbox{d}v>0$, can be simplified as follows:
\begin{equation}
J_u\frac{\partial h}{\partial u}+h\frac{\partial J_u}{\partial u}+J_v\frac{\partial h}{\partial v}+h\frac{\partial J_v}{\partial v}=0
\label{eq2.19}
\end{equation}
Finally, introducing the constitutive equation, equation (\ref{eq2.16}), and the relation between the electric field and electric potential in curvilinear coordinates, equation (\ref{eq2.4}), the continuity equation can be written as follows:
\begin{equation}
k_{vv}\frac{\partial^2 \varphi}{\partial v^2}+2k_{uv}\frac{\partial^2 \varphi}{\partial u\partial v}+k_{uu}\frac{\partial^2 \varphi}{\partial u^2}=0
\label{eq2.20}
\end{equation}

%%%%%%%%%%%%%%%%%%%%%%%%%%%%%%%%%%%%%%%%%%%%%%%%%%%%%%%%%%%%%%%%%%%%%%
\subsection{General solution}
First, let us consider the case for which $u=u_0$ provides the description of $\partial \Omega_{n,0}$. We seek solutions of equation (\ref{eq2.20}) in the form:
\begin{equation}
\varphi\left(u,v\right)=f\left(u+\lambda v\right)
\label{eq2.21}
\end{equation}
where $\lambda$ is a constant and can be complex. Substituting equation (\ref{eq2.21}) into equation (\ref{eq2.20}), one gets: 
\begin{equation}
\left(\lambda^2k_{vv}+2\lambda k_{uv}+k_{uu}\right)f''\left(u+\lambda v\right)=0
\label{eq2.22}
\end{equation}
If one assumes that $f''\neq 0$, then equation (\ref{eq2.22}) provides a characteristic equation for $\lambda$: $\lambda^2k_{vv}+2\lambda k_{uv}+k_{uu}=0$, which is generally satisfied by two complex conjugate solutions. However, thanks to the curvilinear system adopted in this work and the particular arrangement of the fibers, it suffices to consider the case in which $f''=0$ and the electric potential is written as $\varphi\left(u,v\right)=\left(u+\lambda v\right)\zeta$ with $\zeta=$ real constant. This function satisfies the governing equation and the boundary condition at $u=u_0$. In fact, the condition $\textbf{J}\cdot \textbf{n}=J_u=0$ in $\partial \Omega_{n,0}$ reads:
\begin{equation}
\frac {k_{uu}}{h}\zeta+\frac {k_{uv}}{h}\lambda\zeta=0\quad \mbox{for}~u=u_0
\label{eq2.23}
\end{equation}
which is satisfied imposing $\lambda=-k_{uu}/k{uv}$. Accordingly, the following general expression holds for the electric potential:
\begin{equation}
\varphi\left(u,v\right)=\zeta\left(u-\frac{k_{uu}}{k_{uv}}v\right)
\label{eq2.24}
\end{equation}
which, following equation (\ref{eq2.4}), leads to the following current densities in curvilinear components:
\begin{equation}
J_u=0,\quad J_v=\left(-\frac{k_{uv}}{h}+\frac{k_{uu}k_{vv}}{k_{uv}h}\right)\zeta=\frac{\psi}{  \|z'\left(\xi\right)\|}
\label{eq2.25}
\end{equation}
Here, the magnitude of the first derivative of the transformation of coordinates $\|z'\left(\xi\right)\|$ is introduced, $\chi=\left(k_{uu}k_{vv}-k_{uv}^2\right)/k_{uv}$ is a real constant that depends on the electric conductivities and $\psi=\zeta\chi$. It is worth mentioning here that the foregoing curvilinear components of the current density are strikingly identical to the isotropic case whereas the electric potentials and electric fields are different. On the other hand, the electric potential and electric fields take the same expression as the isotropic case when (a) the fiber are aligned  with ($\rho_0=0$) or orthogonal to ($\rho_0=\pi/2$) the field lines or (b) $k_{tt}=k_{mm}$. In such cases, the coupling term $k_{uv}$ goes to zero and the electric potential takes exactly the same form as in the isotropic case. 

The Cartesian components can be found from the curvilinear ones introducing a rotation: $J_x+iJ_y= \exp\left(i\alpha\right)\left(J_u+iJ_v\right)$ with $\alpha$ being the angle between the unit vectors $\textbf{i}$ and $\textbf{e}_u$. Recalling that $\exp\left(i\alpha\right)=\cos \alpha + i \sin \alpha =\sqrt{z'\left(\xi\right)/\bar{z}'\left(\xi\right)}$, where the bar indicates the complex conjugate, the Cartesian components can be written as follows:
\begin{equation}
J_x+iJ_y=i\psi\frac{z'\left(\xi\right)}{\|z'\left(\xi\right)\|^2}
\label{eq2.26}
\end{equation}
As expected, this expression differs from the isotropic case even if the magnitude of the current density remains unchanged. This is because the Cartesian components depend on the anisotropic electric conductivities and $\alpha$, the latter changing continuously in $\Omega$.  

\begin{figure*}
	\center
	\includegraphics[trim=0cm 0cm 0cm 0cm, clip=true,clip=true,width = 1\textwidth]{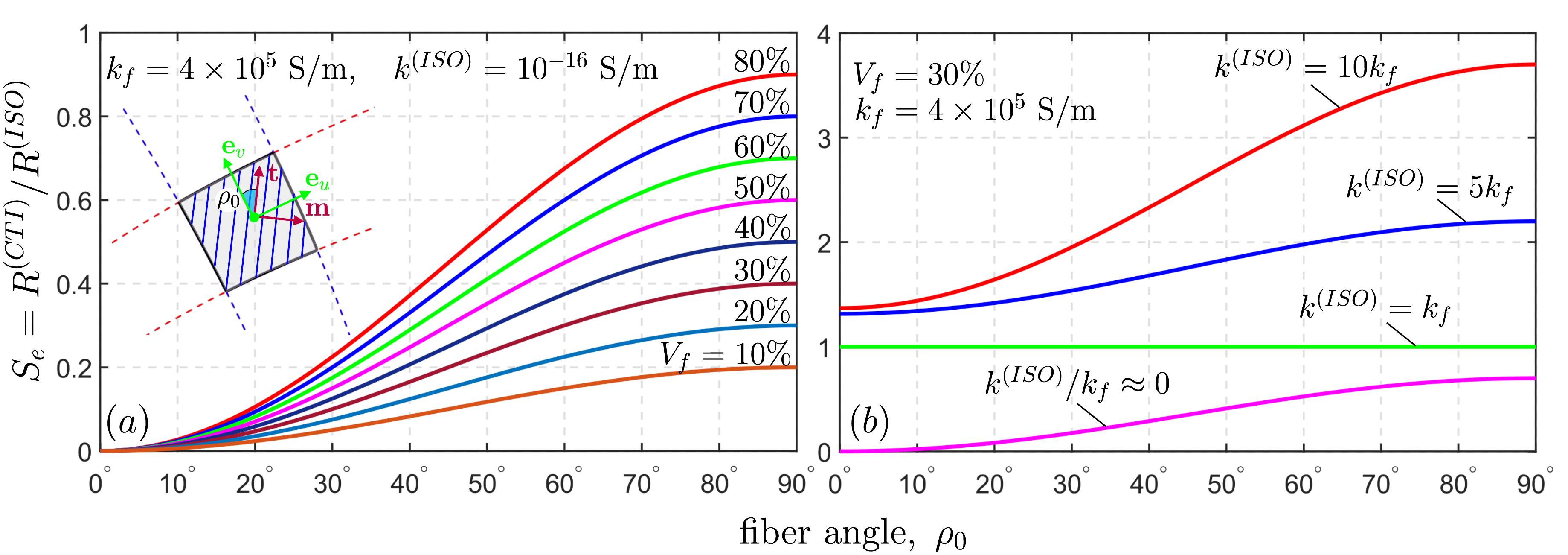}
	\caption{Electrical resistance of plates made of CTI Vs isotropic materials: (a) effect of the volume fraction of conductive fibers; (b) effect of the conductivity of the matrix phase. Equations (\ref{eq2.38})a,b were used to estimate the conductivities in the orthonormal basis $\left(\textbf{t},\textbf{m},\textbf{k}\right)$. The physical properties of the constituents refer to a typical system composed by an epoxy resin reinforced by carbon fibers.}
	\label{f4}
\end{figure*}

The case in which $\partial \Omega_{n,0}$ is described by the condition $v=v_0$ can be solved by means of a similar approach. In such a case, the electric potential is calculated as $\varphi\left(u,v\right)=\zeta\left(v+\lambda u\right)$ where the constant $\lambda$ can be calculated imposing the boundary condition $J_v=0$ when $v=v_0$:
\begin{equation}
\frac{k_{vv}}{h}\zeta+\frac{k_{uv}}{h}\zeta\lambda=0\quad \mbox{for}~~v=v_0
\label{eq2.28}
\end{equation}
According to the foregoing equation, $\lambda=-k_{vv}/k_{uv}$ and the electric potential can be written as follows:
\begin{equation}
\varphi \left(u,v\right)=\zeta\left(v-\frac{k_{vv}}{k_{uv}}u\right)
\label{eq2.29}
\end{equation}
Finally, the curvilinear components of the current density read:
\begin{equation}
J_u=\left(\frac{k_{uv}}{h}-\frac{k_{uu}k_{vv}}{k_{uv}h}\right)\zeta=-\frac{\psi}{\|z'\left(\xi\right)\|},\quad J_v=0
\label{eq2.30}
\end{equation}
where $\psi$ was defined before. Then, the Cartesian components of the current density can be found by means of the following expression:
\begin{equation}
J_u+iJ_v=-\frac{\psi z'\left(\xi\right)}{\|z'\left(\xi\right)\|^2}
\label{eq2.31}
\end{equation}

Finally, it is worth noting that in case the fiber paths are aligned with the field lines ($\rho_0=0$), the coupling term $k_{uv}$ goes to zero. Accordingly, the boundary conditions reduce to the same as the isotropic case and the governing equation, equation (\ref{eq2.20}), simplifies to:
\begin{equation}
k_{vv}\frac{\partial^2 \varphi}{\partial v^2}+k_{uu}\frac{\partial^2 \varphi}{\partial u^2}=0
\label{eq2.32}
\end{equation}
Notwithstanding the presence of the terms $k_{uu}$ and $k_{vv}$, this equation has the same solution as the isotropic case: 
\begin{equation}
\varphi\left(u,v\right)= \zeta v
\label{eq2.32bis}
\end{equation}
for $u=u_0$ describing $\partial \Omega_{n,0}$ and:
\begin{equation}
\varphi\left(u,v\right)= \zeta u
\label{eq2.32bisbis}
\end{equation}
for $v=v_0$ describing $\partial \Omega_{n,0}$. Accordingly, the current density can be calculated using equations (\ref{eq2.9bis}), (\ref{eq2.10}), (\ref{eq2.9bisbis}) and (\ref{eq2.11}) provided that the proper conductivities $k_{vv}$ and $k_{uu}$ are used instead of $k$ in the calculation of $\psi$. 

\subsection{Optimal fiber path for conductivity and calculation of the electrical resistance}
Leveraging the foregoing solutions, it is possible to investigate the reduction of electrical resistance enabled by the curvilinear fibers compared to the isotropic case. Towards this goal, let us focus on the case in which $\forall u \in \partial \Omega_{n,0}, u=u_0$ and consider an infinitesimal flux tube. As was shown in the foregoing sections, the flux tube is defined by curves $u=const$ either when the material is isotropic or when it exhibits curvilinear transverse isotropy since $J_v$ is the only nonzero curvilinear component. Assuming that the same potential difference $\Delta \varphi_0$ is applied at the remote extremities of the flux tube, the electric current can be expressed as $\mbox{d}I=J_v h\mbox{d}u\mbox{d}z$ where the current density can be calculated by means of the proposed formulation and it is considered uniform through the thickness. Finally, the electrical resistance of the domain can be obtained by integration over all the flux tubes, leading to the following expression:
\begin{equation}
\begin{aligned}
R^{\left(ISO\right)}&=\frac{\Delta \varphi_0}{I} \quad \mbox{where} \\ ~~I&=\int_0^T{\int_{u_0}^{u_L}{J_v^{\infty}\left(u\right)h\mbox{d}u}\mbox{d}z}=\psi^{(ISO)} T\left(u_L-u_0\right)
\label{eq2.33}
\end{aligned}
\end{equation}
for the isotropic case and:
\begin{equation}
\begin{aligned}
R^{\left(CTI\right)}&=\frac{\Delta \varphi_0}{I}\quad \mbox{where} \\ ~~I&=\int_0^T{\int_{u_0}^{u_L}{J_v^{\infty}\left(u\right)h\mbox{d}u}\mbox{d}z}=\psi^{(CTI)} T\left(u_L-u_0\right)
\label{eq2.34}
\end{aligned}
\end{equation}
for the case of curvilinear transverse isotropy. In the foregoing expressions, $T=$ plate thickness and $\psi^{(ISO)}$ and $\psi^{(CTI)}$ are real constants defined in equations (\ref{eq2.9bis}) and (\ref{eq2.25}) respectively. Thanks to equations (\ref{eq2.33}) and (\ref{eq2.34}), one can easily calculate the resistance. 

A particularly important conclusion that can be drawn from the foregoing results is that, since the field lines for the isotropic and transversely isotropic case are identical, the resistivity differs only for a multiplying constant dependent on the fiber angle, $\rho_0$, and anisotropic electric conductivities $k_{tt}$ and $k_{mm}$. This allows a very simple description of the ratio between the electrical resistance of the transversely isotropic body and the isotropic one: $S_e\left(\rho_0, k_{tt}, k_{mm}\right)=R^{\left(CTI\right)}/R^{\left(ISO\right)}$. In fact, assuming that the same potential differential $\Delta \varphi_0$ is applied to the two bodies then one gets $\Delta \varphi_0=2\zeta^{(ISO)}v_0=2\zeta^{(CTI)}v_0 k_{uu}/k_{uv}$ with $v_0$ representing the value of the curvilinear coordinate $v$ where the potential is imposed. Accordingly, $\zeta^{(CTI)}=\zeta^{(ISO)}k_{uv}/k_{uu}$ and the constant $\psi$ can be written as follows:
\begin{equation}
\psi^{(CTI)}=\zeta^{\left(ISO\right)}\frac{k_{uu}k_{vv}-k_{uv}^2}{k_{uu}}=\psi^{\left(ISO\right)}\frac{k_{uu}k_{vv}-k_{uv}^2}{k_{uu}k^{(ISO)}}
\label{eq2.35}
\end{equation}
Leveraging equations (\ref{eq2.33}), (\ref{eq2.34}) and, (\ref{eq2.35}) and using its definition, $S_e$ can finally be calculated as follows:
\begin{equation}
\begin{aligned}
S_e\left(\rho_0,k_{tt},k_{mm}\right)&=\frac{R^{\left(CTI\right)}}{R^{\left(ISO\right)}}=\frac{k_{uu}k^{(ISO)}}{k_{uu}k_{vv}-k_{uv}^2} \\ &=k^{(ISO)}\left[\frac{\cos^2\left(\rho_0\right)}{k_{tt}}+\frac{\sin^2\left(\rho_0\right)}{k_{mm}}\right]
\label{eq2.36}
\end{aligned}
\end{equation}
where $k^{(ISO)}$ represents the electrical conductivity of the isotropic material. It is worth mentioning that equation (\ref{eq2.36}) is very useful for design since it provides a very simple expression to estimate the electrical resistance reduction induced by the addition of the conductive fibers. Further, it simplifies significantly the computational modeling of materials featuring curvilinear transverse isotropy, in case a numerical simulation is preferred over a closed-form solution (which may be the case when the geometry cannot be described by a relatively simple conformal map). In fact, if the angle $\rho_0$ as well as the conductivities $k_{tt}$ and $k_{mm}$ are known, one can analyze the isotropic medium using standard Finite Element Analysis (FEA) and then use equation (\ref{eq2.36}) to calculate the resistance for the curvilinear, transversely isotropic case. Even the current densities can be calculated from the isotropic case simply leveraging the link between the isotropic constants and $\psi$ provided in equation (\ref{eq2.35}) and by substituting into equation (\ref{eq2.25}). The advantage is that, this way, the fibers do not need to be considered explicitly and the analysis can be conducted in any standard FE software without the need for specialized subroutines. 

\begin{figure*}
	\center
	\includegraphics[trim=0cm 0cm 0cm 0cm, clip=true,clip=true,width = 0.8\textwidth]{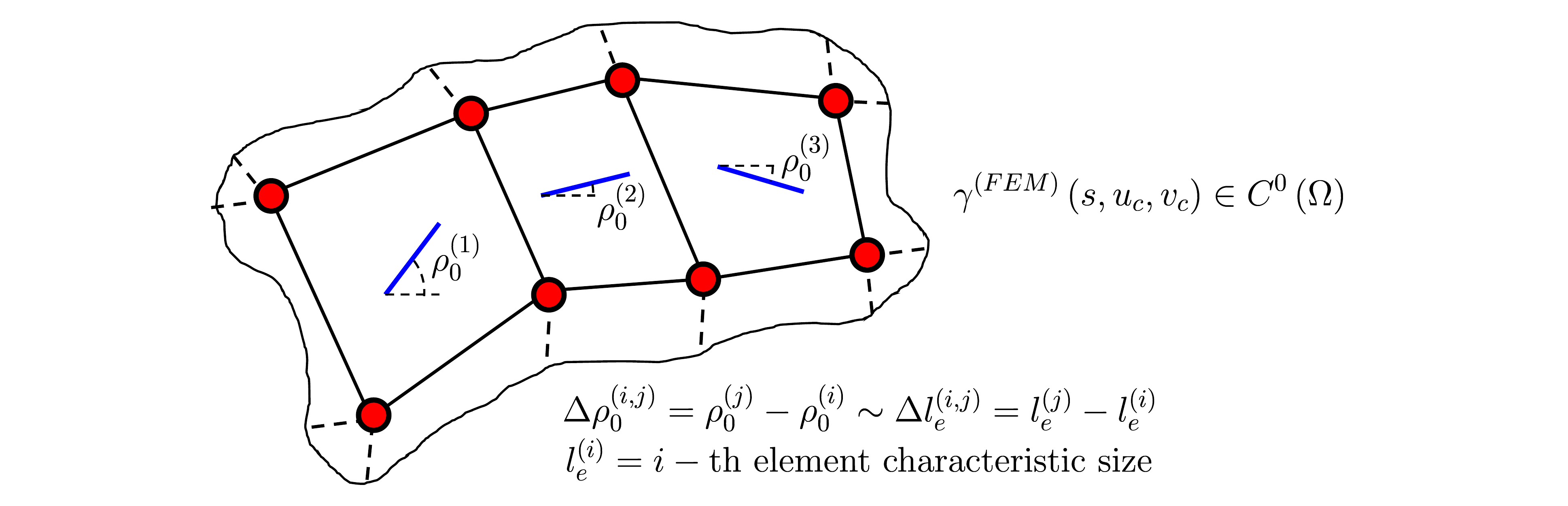}
	\caption{Schematic representation of a patch of $4-$node, isoparametric, quadrilateral elements with material orientation assigned to each integration point by the user subroutine \textit{ORIENT} in ABAQUS/standard \cite{ABAQUS}. As can be noted, the curves describing the fiber paths are only $C^0\left(\Omega\right)$.}
	\label{f5}
\end{figure*}

equation (\ref{eq2.36}) enables the identification of the fiber orientation leading to the minimum $S_e$, i.e. the maximum reduction of electrical resistance. Considering that the fibers are conductive and that they act in parallel coupling with the matrix along $\textbf{t}$ and in series coupling along $\textbf{m}$, it is reasonable to assume that $k_{tt}\geq k_{mm} \geq k^{(ISO)}>0$. In the particular, unlikely case in which $k_{tt}=k_{mm}=k$, $S_e$ is minimized by the condition that $k$ is maximum, independently on the value taken by $\rho_0 \in [0,\pi)$. When $k_{tt}\neq k_{mm}$, then the configuration providing the minimum electrical resistance is always the one in which the fibers are aligned with the field lines, i.e. $\rho_0=0$. In such a case:
\begin{equation}
S_e\left(0,k_{tt},k_{mm}\right)=\frac{k^{(ISO)}}{k_{tt}}
\label{eq2.37}
\end{equation}
which depends only on the ratio between the conductivity of the matrix and the conductivity in the fiber direction. The latter conductivity depends on the volume fraction of fibers, $V_f$, the conductivity of the fibers $k_f$ and of the matrix $k_{mat}=k^{(ISO)}$. Considering the coupling between fibers and matrix, a first order approximation of $k_{tt}$ and $k_{mm}$ can be obtained as follows:
\begin{equation}
\begin{aligned}
k_{tt}&=V_f k_f + \left(1-V_f\right) k_{mat},\\ 
k_{mm}&=k_f k_{mat}\left[V_f k_m+\left(1-V_f\right)k_f\right]^{-1}
\label{eq2.38}
\end{aligned}
\end{equation}
In case that a more refined estimate is needed, several micromechanical models can be found in the literature \cite[e.g.]{Eshelby57, MoTa73, SevKac09, Mura13, NemHori13}.   

A summary of the foregoing conclusions is provided in Figure \ref{f4}a which shows the effect of the curvilinear fiber angle on the electrical resistance for various volume fractions of fibers assuming a fiber conductivity $k_f=4\times 10^5$ S/m and a matrix conductivity $k^{(ISO)}=10^{-16}$ S/m. These values correspond to a typical epoxy system reinforced by carbon fibers. As can be noted, the electrical resistance decreases dramatically decreasing the angle $\rho_0$, regardless of the volume fraction. On the other hand, a higher amount of conductive fibers leads to lower electrical resistance for a given angle. The importance of the fiber angle is also confirmed by Figure \ref{f4}b which reports the electrical resistance against the fiber angle for various conductivities of the matrix, $k^{(ISO)}$. As can be noted, the minimum resistance is always achieved when $\rho_0=0$ and the fibers and the matrix are in parallel coupling along the field lines, regardless of the conductivity of the matrix. In contrast,  when the conductivity of the matrix is equal to the one of the fibers, the electrical resistance becomes independent of the fiber angle. 

Finally, it is worth mentioning that the case in which $\forall v \in \partial \Omega_{n,0}, v=v_0$ provides the same expression for $S_e$ as reported in equation (\ref{eq2.36}).

\section{Results and discussion}
In this section, the theoretical framework is applied to several examples of notched plates. In each case, the explicit equations of the the fiber paths minimizing the electrical resistance are provided along with the expressions for electric potential and the current density components. The results are validated by comparing the theoretical predictions to a large bulk of Finite Element electrostatic analyses conducted in ABAQUS $6.14$ \cite{ABAQUS}. The electric potential and current density were investigated leveraging a $2$D mesh of DC$2$D$8$, $8-$node isoparametric quadrilateral elements with anisotropic linear electrostatic constitutive laws by means of the implicit solver ABAQUS/standard \cite{ABAQUS}. The user subroutine \textit{ORIENT} \cite{ABAQUS} was used to prescribe a material orientation following the assigned curvilinear paths. It is worth mentioning that \textit{ORIENT} assigns a uniform fiber orientation to each element integration point. Accordingly, the description of the fiber paths obtainable by this method, $\gamma^{(FEM)}\left(s,u_c,v_c\right)$, is only $C^0\left(\Omega\right)$ (see Figure \ref{f5}). For this reason, particular care was devoted in verifying the numerical convergence of all the cases investigated, particularly the ones involving sharp corners leading to singular current density fields. A simulation was assumed to converge  when  further  mesh refinements would lead to changes in the maximum current density values lower  than $0.01\%$. This convergence criterion led to meshes featuring elements typically one order of magnitude smaller compared to the isotropic case due to the poor geometrical description of the fibers. To remedy this issue, a new Finite Element based on Isogeometric Analysis (IGA) \cite{BasHug09} enabling a smooth description of the fibers is under development by the authors. Its description is beyond the scope of the present work.    
\begin{figure*}
	\center
	\includegraphics[trim=0cm 0cm 0cm 0cm, clip=true,clip=true,width = 1\textwidth]{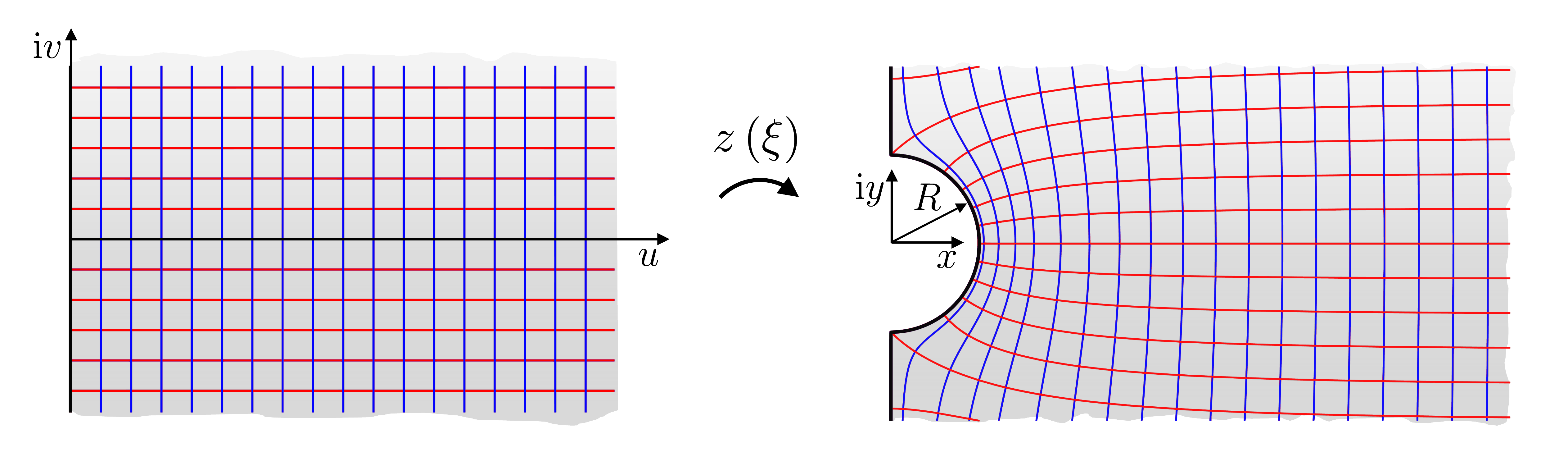}
	\caption{Schematic representation of the conformal map provided by equation (\ref{eq2.80}), transforming the right half-plane into a semi-infinite plate featuring a semi-circular notch of radius $R$}
	\label{f6}
\end{figure*}

\subsection{Semicircular notch}
To exemplify the application of the foregoing theoretical framework, let us consider the following conformal transformation:
\begin{equation}
z\left(\xi\right)=R\left(\frac{\xi}{2}+\frac{1}{2}\sqrt{\xi^2+4}\right)
\label{eq2.80}
\end{equation}
This mapping transforms the right curvilinear half-plane, $u\ge0$, into an semi-infinite plate featuring a semi-circular notch of radius $R$ (Figure \ref{f6}). The boundary, which is considered isolated, $\textbf{J}\cdot \textbf{n}=0$, is described by the condition $u=0$ whereas the bisector of the notch is described by the condition $v=0$. It is worth mentioning here that, even if the foregoing map considers a semi-infinite medium, it can also be applied to finite bodies provided that the domain is large enough so that the effects of the boundaries are negligible. Further, this transformation can always be used to describe the electric potential and current density sufficiently close to the notch, no matter the size of the domain. In such a case, however, the constant $\psi$ required for the calculation of the current density components (see equation \ref{eq2.25}) must be estimated by e.g. FEA. Alternatively, in case of finite bodies, one can always find the exact solution of equation (\ref{eq2.20}) by assuming $f'',\lambda \neq 0$ and leveraging a Laurent series. However, this latter approach would prevent the achievement of a closed-form solution.

\subsubsection{Optimal fiber paths}
\begin{figure*}
	\center
	\includegraphics[trim=0cm 0cm 0cm 0cm, clip=true,clip=true,width = 1\textwidth]{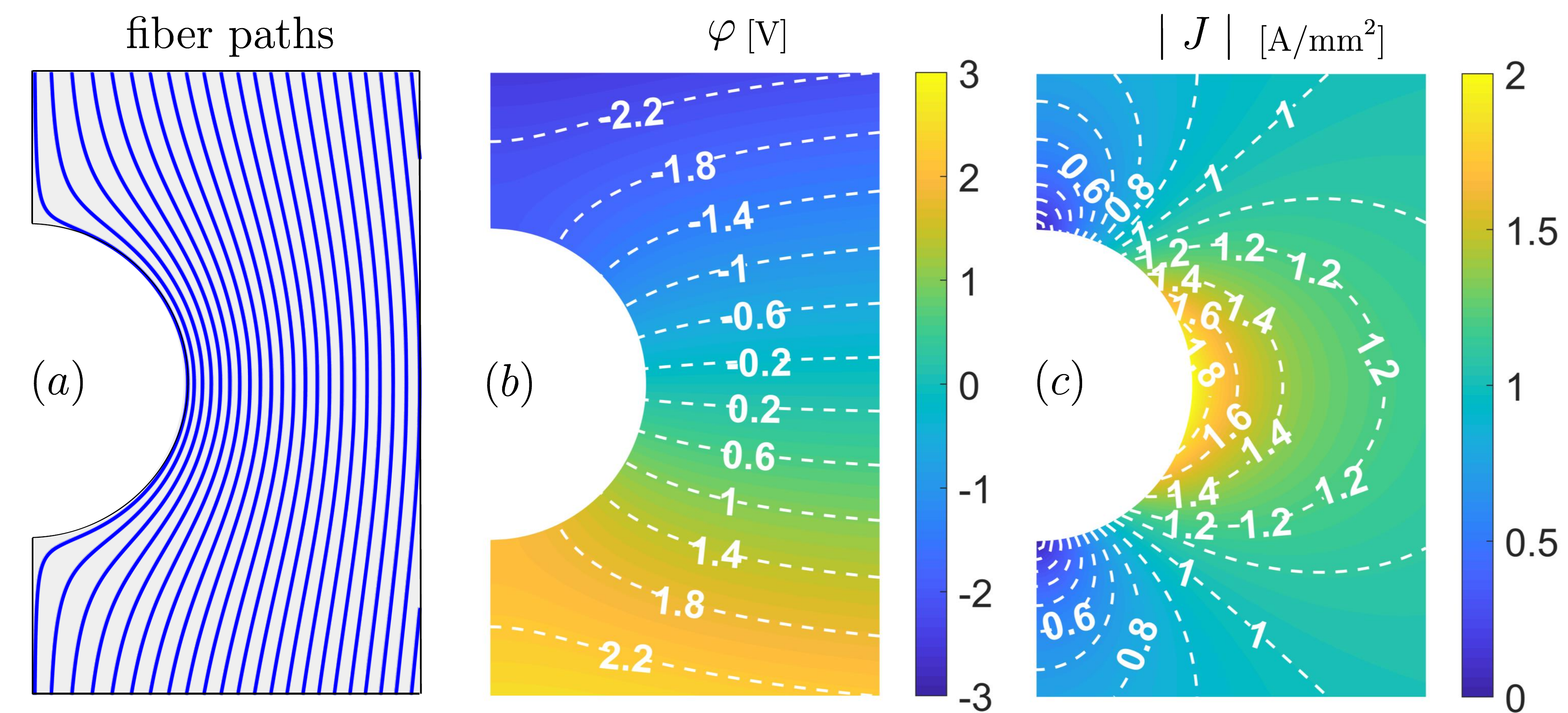}
	\caption{Solution of the electrostatic problem for a semi-infinite body featuring curvilinear transverse isotropy weakened by semi-circular notch: (a) fiber paths of minimum electrical resistance; (b) electric potential field; (c) magnitude of the electric current density. The notch radius is $R=1$ mm.}
	\label{f7}
\end{figure*}

In Section \ref{CTI}, it was shown that the fiber paths, $\gamma\left(s,u_c,v_c\right)$, providing the minimum electrical resistance are aligned with the field lines. Leveraging equations (\ref{eq2.13}a,b) and (\ref{eq2.80}) and imposing $\rho_0=0$, the paths can be calculated easily as follows:

\begin{subnumcases}{\gamma\left(s,u_c,v_c\right):}
	x\left(s,u_c,v_c\right)=\frac{R u_c}{2}+\frac{R}{2}\left\{\left[u_c^2-\left(s+v_c\right)^2\right]^2+4u_c^2\left(s+v_c\right)^2\right\}^{\frac{1}{4}}\cos\frac{\eta}{2}  \\
	y\left(s,u_c,v_c\right)=\frac{R}{2}\left\{\left[u_c^2-\left(s+v_c\right)^2\right]^2+4u_c^2\left(s+v_c\right)^2\right\}^{\frac{1}{4}}\sin\frac{\eta}{2}  
\end{subnumcases}\label{eq2.81}

where $\eta=\Arg \left[u_c^2-\left(s+v_c\right)^2+2\mbox{i}u_c\left(s+v_c\right)\right]$. A schematic representation of the foregoing fiber paths for $R=1$ mm can be found in Figure \ref{f7}a.

\subsubsection{Electric potential}
The electric potential can be obtained easily by means of equations (\ref{eq2.32bis}) and (\ref{eq2.80}). In fact, noting that $\xi\left(z\right)=z/R-R/z$, the curvilinear coordinate $v$ can be determined easily, leading to the following result in Cartesian coordinates:
\begin{equation}
\varphi\left(x,y\right)=\zeta \left(\frac{\sqrt{x^2+y^2}}{R}+\frac{R}{\sqrt{x^2+y^2}}\right)\sin\left[\Arg\left(x+\mbox{i}y\right)\right]
\label{eq2.82}
\end{equation}
Alternatively, equation (\ref{eq2.82}) can be written in polar coordinates as follows:
\begin{equation}
\varphi\left(r,\theta\right)=\zeta \left(\frac{r}{R}+\frac{R}{r}\right)\sin\theta
\label{eq2.83}
\end{equation}
The electric potential for $R=1$ and $\zeta=-1$ V/mm is provided in Figure \ref{f7}b.

\subsubsection{Current density}
\begin{figure*}
	\center
	\includegraphics[trim=0cm 0cm 0cm 0cm, clip=true,clip=true,width = 1\textwidth]{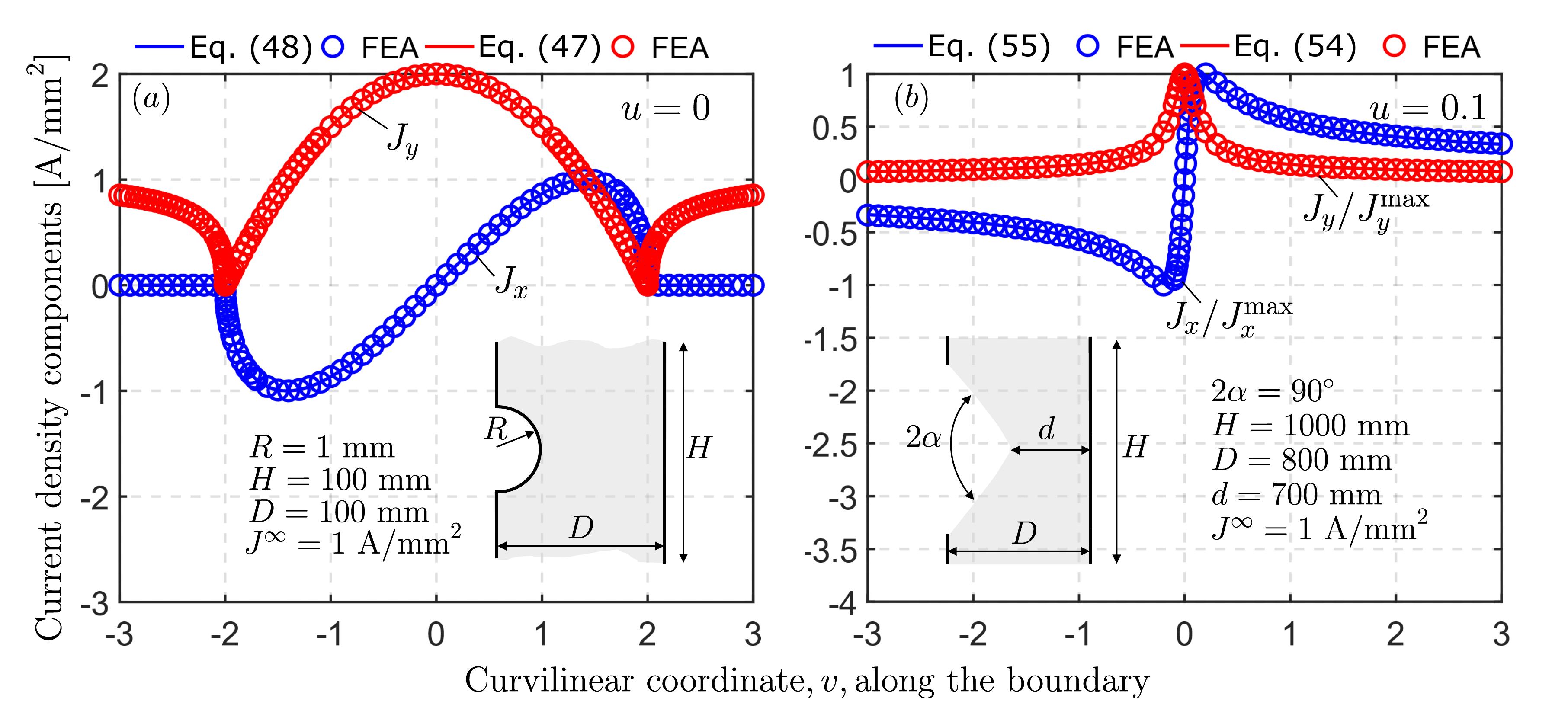}
	\caption{Comparison between the theoretical solution and FEA: (a) Cartesian components of the current density along the boundary of a plate featuring a semicircular notch; (b) normalized Cartesian components of the current density along the boundary of a plate featuring a hyperbolic notch of opening $2\alpha=90^{\circ}$ and tip radius $\rho=95$ $\mu$m.}
	\label{f8}
\end{figure*}

Finally, the current density components can be calculated by means of equation (\ref{eq2.10}):
\begin{equation}
J_y+\mbox{i}J_x=\frac{\psi}{R}\left[1+\left(\frac{R}{z}\right)^2\right]
\label{eq2.84}
\end{equation}
which can be used to calculate the Cartesian components of the current density in polar coordinates:
\begin{equation}
J_y\left(r,\theta\right)=\frac{\psi}{R}\left[1+\left(\frac{R}{r}\right)^2\cos 2\theta\right]
\label{eq2.85}
\end{equation}
\begin{equation}
J_x\left(r,\theta\right)=-\frac{\psi}{R}\left(\frac{R}{r}\right)^2\sin 2\theta
\label{eq2.86}
\end{equation}
Figure \ref{f7}c shows a contour plot of the magnitude of the current density for $R=1$ and $\psi=1$ A/mm which corresponds to the case of a unitary remote positive current, i.e. $J_y^{\infty}=1$ A/mm$^2$. Figure \ref{f8}a shows a comparison between the current density components along the boundary ($u=0$) calculated by means of equations (\ref{eq2.85}) and (\ref{eq2.86}) and FEA. In the numerical analysis, a plate of height $H=100$ mm, width $D=100$ mm and notch radius $R=1$ mm was considered. It can be seen that theoretical and numerical results are in perfect agreement, with differences that are always lower than $0.2\%$. It should also be noted that, as expected, the notch leads to a concentration of current density. Following equations (\ref{eq2.85}) and (\ref{eq2.86}), the maximum current occurs at the notch tip with the maximum concentration factor $K=J^{max}/J_y^{\infty}=2$.

\subsection{Deep hyperbolic notch in an infinite plate}
Let us consider the mapping represented in Figure \ref{f1}b \cite{Neu58,Neu85}:
\begin{equation}
z\left(\xi\right)=\xi^q
\label{eq2.39}
\end{equation}
where $u=u_0$ represents the condition for the boundary of a deep hyperbolic notch, $q$ is a parameter linked to the notch opening angle, $2\alpha$, by the expression $q=\left(2\pi-2\alpha\right)/\pi$ with $q\in[1,2]$. Notwithstanding its simplicity, the conformal map described by equation (\ref{eq2.39}) is rather general since it can describe infinite notches of any opening angle and radius of curvature at the tip. In fact, the radius can be controlled by the expression $\rho=u_0^q q/(q-1)$ which shows that the particular case $u_0=0$ describes a sharp notch of opening $2\alpha$. The case of a crack, i.e. $\rho=0$ and $2\alpha=0$, can be obtained imposing $u_0=0$ and $q=2$ \cite{Neu58, Neu85}. When $q=3/2$ and $u_0=0.1$, the opening angle is $90^{\circ}$ with a tip radius of $95$ $\mu$m as shown in Figure \ref{f9}.    
\subsubsection{Optimal fiber paths}
The field lines close to the notch can be calculated easily leveraging equation (\ref{eq2.13}) which, using the composed function $z\left(s\right)=\left(z \circ\xi\right)\left(s\right)$, leads to the following expression in Cartesian coordinates:

\begin{subnumcases}{\gamma\left(s,u_c,v_c\right):}
	x\left(s,u_c,v_c\right)=\left[u_c^2+\left(v_c+s\right)^2\right]^{q/2}\cos\left\{ q \Arg \left[u_c+i\left(v_c+s\right)\right]\right\}\label{eq2.401}  \\
	y\left(s,u_c,v_c\right)=\left[u_c^2+\left(v_c+s\right)^2\right]^{q/2}\sin\left\{ q \Arg \left[u_c+i\left(v_c+s\right)\right]\right\}\label{eq2.402}  
\end{subnumcases}

Here $s\in\left[0,\infty\right)$, $u_c=u\left(0\right)$, $v_c=v\left(0\right)$ and $\rho_0$ was set to zero. A schematic representation of the field lines, equations (\ref{eq2.401}) and (\ref{eq2.402}), is provided in Figure \ref{f9}a. These curves represent the fiber paths minimizing the electrical resistance of the body.
\subsubsection{Electric potential}
\begin{figure*}
	\center
	\includegraphics[trim=0cm 0cm 0cm 0cm, clip=true,clip=true,width = 1\textwidth]{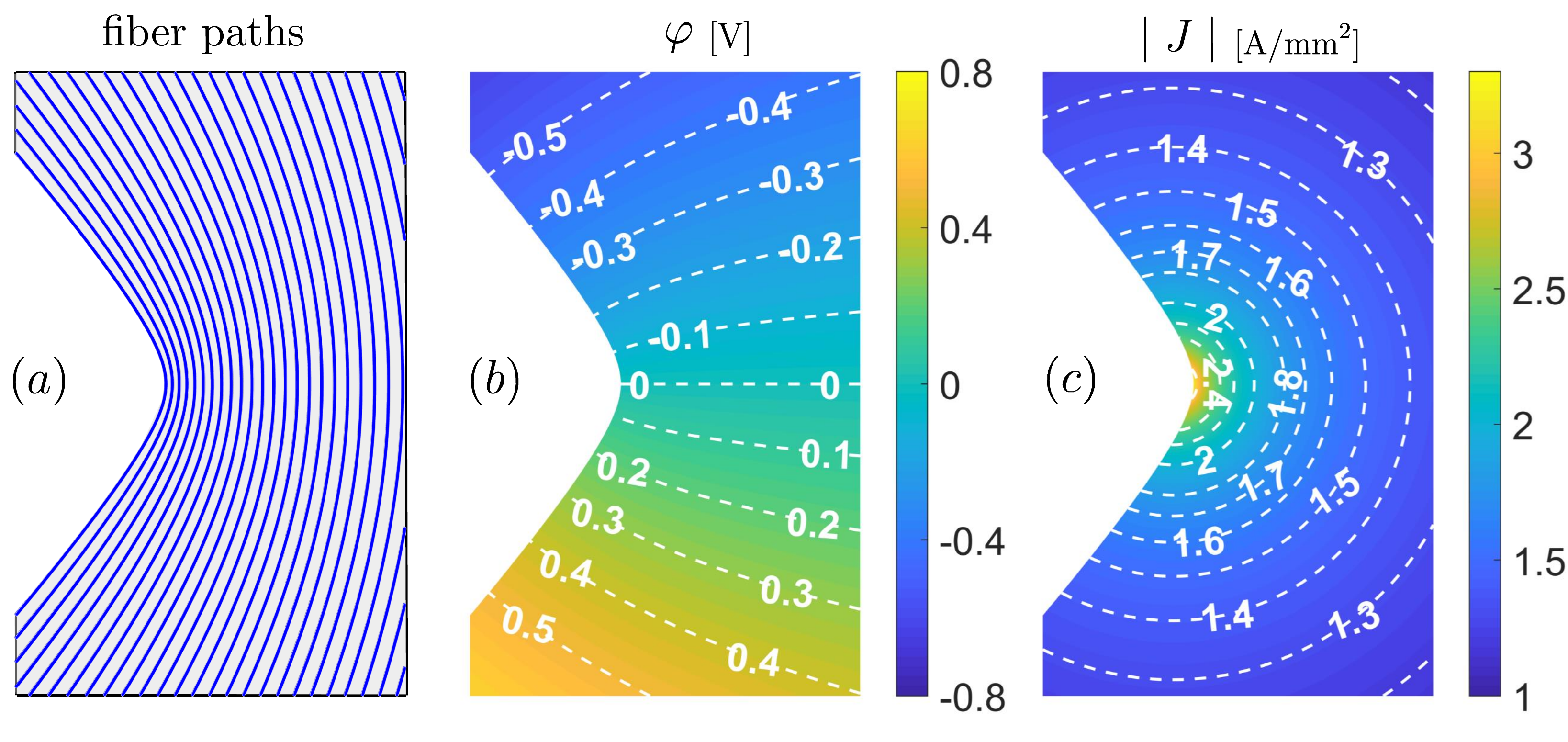}
	\caption{Solution of the electrostatic problem for an infinite body featuring curvilinear transverse isotropy weakened by a deep hyperbolic notch: (a) fiber path of minimum electrical resistance; (b) electric potential field; (c) magnitude of the electric current density. The notch is described by $z\left(\xi\right)=\xi^q$ with $q=3/2$, $u_0=0.1$ corresponding to an opening angle $2\alpha=90^{\circ}$ and a notch tip radius $\rho=95$ $\mu$m.}
	\label{f9}
\end{figure*}

Once the conformal mapping describing the domain is identified, the electric potential can be calculated easily in Cartesian or polar coordinates leveraging equation (\ref{eq2.9}): 
\begin{equation}
\varphi\left(x,y\right)=\zeta\left(x^2+y^2\right)^{1/2q}\sin\left[\frac{1}{q}\Arg \left(x+iy\right)\right]
\label{eq2.40bisbisbis}
\end{equation}
and
\begin{equation}
\varphi\left(x,y\right)=\zeta r^{1/q}\sin\left(\frac{\theta}{q}\right)
\label{eq2.40bis}
\end{equation}
It is worth mentioning that, in this particular case, the electric potential can be calculated but for the multiplying constant $\zeta$ since both the notch and the domain are infinite. This constant can be calculated analytically or numerically by analyzing a finite plate with a deep hyperbolic notch and a sufficiently large ligament. Figure \ref{f9}b shows the distribution of electric potential for $\zeta=-1$ V/mm$^{1/q}$.

\subsubsection{Current density}
The current density can be calculated from the first derivative of the Neuber's map, equation(\ref{eq2.39}), leading to the following equation:
\begin{equation}
J_y+iJ_x=\frac{\psi} {q} z^{1/q-1}
\label{eq2.41}
\end{equation}
which can be used to calculate the Cartesian components of the current density in polar coordinates:
\begin{equation}
J_y\left(r,\theta\right)=\frac{\psi} {q} r^{1/q-1}\cos\left[\left(\frac{1}{q}-1\right)\theta\right]
\label{eq2.42}
\end{equation}
\begin{equation}
J_x\left(r,\theta\right)=\frac{\psi} {q} r^{1/q-1}\sin\left[\left(\frac{1}{q}-1\right)\theta\right]
\label{eq2.43bis}
\end{equation}
Figure \ref{f9}c shows the distribution of the magnitude of the current density for $\psi/q=1$ A/mm$^{1/q+1}$ whereas Figure \ref{f8}b shows a comparison between the normalized current density components along the notch boundary ($u=0.1$) calculated by means of equations (\ref{eq2.42}) and (\ref{eq2.43bis}) and FEA. In the numerical analysis, a plate of height $H=1000$ mm, width $D=800$ mm, ligament length $d=700$ mm, notch opening $2\alpha=90^{\circ}$ and tip radius $\rho=95$ $\mu$m was considered. It can be seen that theoretical and numerical results are in perfect agreement, with differences that are always lower than $0.25\%$.

\subsection{Sharp V-notch of finite depth}
The link between the electrostatic solution and the conformal map used to describe the domain makes methodologies such as the Schwarz--Christoffel transformation \cite{DriTre02}, that allows the geometrical construction of the map for any polygonal domain, particularly convenient. In general terms, the Schwarz--Christoffel Mapping (SCM) can be written as \cite{DriTre02}:
\begin{equation}
z\left(\xi\right)=z\left(\xi_0\right)+C\int_{\xi_0}^{\xi}{\prod_{j=1}^n{\frac{\mbox{d}w}{\left(w-a_j\right)^{1-\alpha_j/\pi}}}}
\label{eq2.43}
\end{equation}
where $C$ is a constant and $z\left(\xi\right)$ maps the real axis, $v\left(x,y\right)=0$, to the edges of a polygon with $n$ sides and interior angles $\alpha_j$ with $a_j$ being the vertices of the polygon in the Argand plane.
\begin{figure*}
	\center
	\includegraphics[trim=0cm 0cm 0cm 0cm, clip=true,clip=true,width = 1\textwidth]{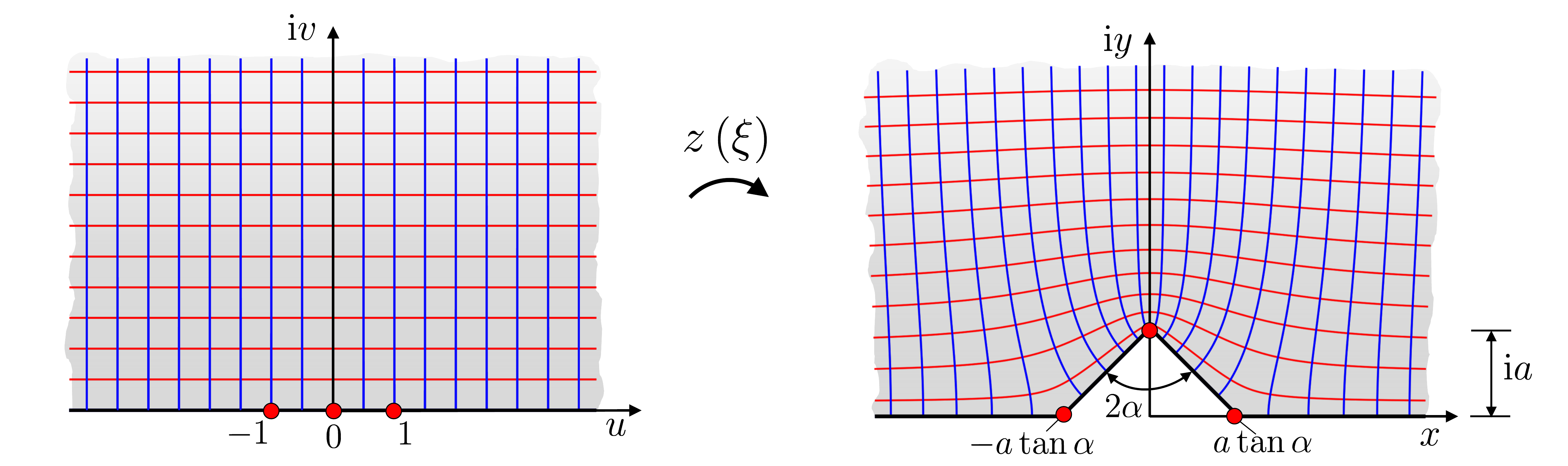}
	\caption{Schematic representation of the conformal map provided by equation (\ref{eq2.44}), transforming the half-plane, $v\ge0$, into a semi-infinite plate featuring a finite V-notch of depth $a$ and opening $2\alpha$.}
	\label{f10}
\end{figure*}
Let us consider the case of a finite, sharp V-notch in an infinite solid subjected to a remote current $J^{\infty}$ or a remote potential differential $\Delta \varphi_0$. Thanks to the SCM, the conformal map describing the notch profile and the corresponding first derivative are:
\begin{equation}
z'\left(\xi\right)=A\frac{\xi^{1-2\alpha/\pi}}{\left(\xi^2-1\right)^{1/2-\alpha/\pi}},\quad z\left(\xi\right)=B+\int^{\xi}{z'\left(w\right)\mbox{d}w}
\label{eq2.44}
\end{equation}
where $A$ and $B$ are complex constants. Solving the integral in the second of equation (\ref{eq2.44}) and imposing the following conditions:
\begin{subnumcases}{}
z\left(\xi=0\right)=ia \\
y\left(\xi=1\right)=a\tan \alpha  
\end{subnumcases}\label{eq2.45}
one obtains:
\begin{equation}
A=\frac{a\sqrt{\pi}}{\cos \alpha \Gamma\left(1-\frac{\alpha}{\pi}\right)\Gamma\left(1+\frac{\alpha}{\pi}\right)},\quad B=ia
\label{eq2.46}
\end{equation}
and:
\begin{equation}
\begin{aligned}
z\left(\xi\right)&=B+A\frac{\pi\xi^{2\left(1-\frac{\alpha}{\pi}\right)}}{2\left(\pi-\alpha\right)}\left(\frac{1-\xi^2}{\xi^2-1}\right)^{\frac{\alpha}{\pi}-\frac{1}{2}}{}H(\zeta) \\
&\mbox{where} \quad H(\zeta)=_2F_1\left(\frac{1}{2}-\frac{\alpha}{\pi},1-\frac{\alpha}{\pi},2-\frac{\alpha}{\pi},\xi^2\right)
\label{eq2.47}
\end{aligned}
\end{equation}
In the foregoing equations, $a$ is the depth of the notch, $2\alpha$ is the opening angle, $\Gamma\left(t\right)=\int_{0}^{\infty}{x^{t-1}\exp{\left(-x\right)}\mbox{d}x}$ is the gamma function \cite{AbrSte72} and, ${}_2F_1\left(a,b,c,z\right)=\sum_{k=0}^{\infty}{\left(a\right)_k\left(b\right)_k/\left(c\right)_k z^k/k!}$ is the Gaussian hypergeometric function \cite[e.g.]{Yos97, AndAskRoy99}. A schematic representation of the conformal transformation is provided in Figure \ref{f10}.

\begin{figure*}
	\center
	\includegraphics[trim=0cm 0cm 0cm 0cm, clip=true,clip=true,width = 1\textwidth]{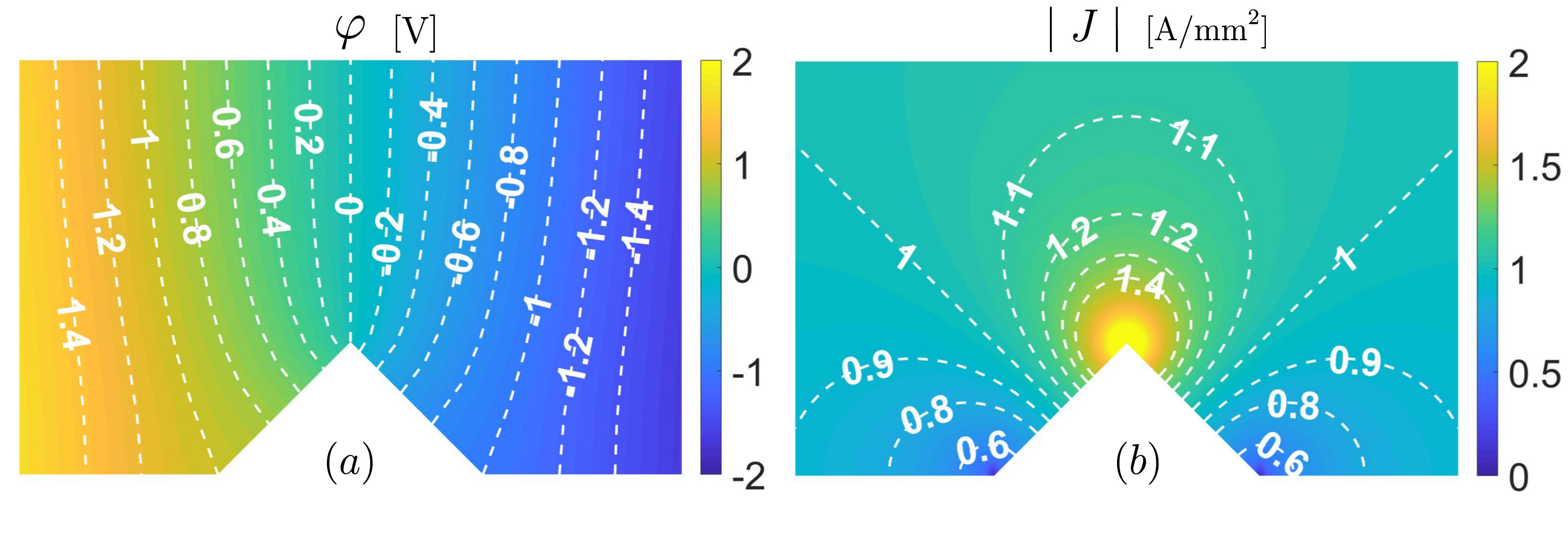}
	\caption{Electrostatic solution for an infinite transversely isotropic body featuring a sharp V-notch of finite depth $a=1$ and opening angle $2\alpha=90^{\circ}$. (a) Electric potential distribution; (b) magnitude of the current density.}
	\label{f11}
\end{figure*}

\subsubsection{Optimal fiber paths}
In this case, the boundary is described by the condition $v=v_0=0$. According to Section \ref{remarks}, the optimal fiber paths can be found by means of equation (\ref{eq2.14}) imposing $x\left(s,u_c,v_c\right)=\operatorname{Re}\left[\left(z \circ\xi\right)\left(s,u_c,v_c\right)\right]$ and $y\left(s,u_c,v_c\right)=\operatorname{Im}\left[\left(z \circ\xi\right)\left(s,u_c,v_c\right)\right]$. The dashed lines in Figure \ref{f2}a,b show a schematic representation of the fiber paths that minimize the electrical resistance of the body, coinciding with the field lines, both in the physical and transformed domain. The continuous lines represent curves that are rotated of $45^{\circ}$ with respect to the field lines.
\subsubsection{Electric potential}
The electric potential can be calculated recalling that, in case $v=v_0$ describes the isolated boundary $\partial \Omega_{n,0}$, the potential is $\varphi\left(u,v\right)=\zeta u$. Since the solution in Cartesian coordinates cannot be found easily in closed-form owed to the complexity of equation (\ref{eq2.47}), only the implicit description of the potential is provided as follows:
\begin{equation}
\varphi\left(x,y\right)=\zeta\operatorname{Re}\left[\xi\left(z\right)\right]
\label{eq2.47bis}
\end{equation}
Figure \ref{f11}a show the contour plot of the electric potential for $\zeta=1$ V. 

\subsubsection{Current density}
\begin{figure*}
	\center
	\includegraphics[trim=0cm 0cm 0cm 0cm, clip=true,clip=true,width = 1\textwidth]{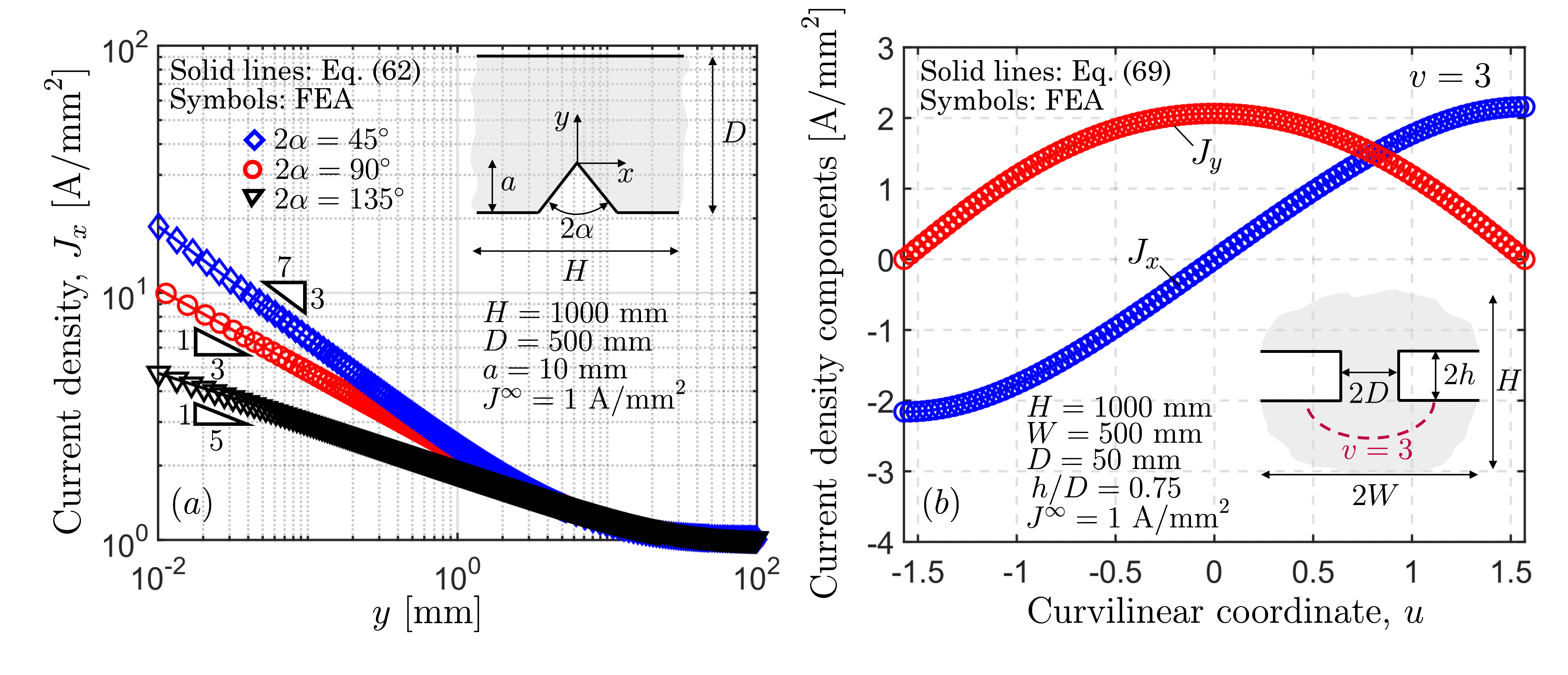}
	\caption{Comparison between the theoretical solution and FEA: (a) double-logarithmic plot of the current density component, $J_x$, along the bisector of a finite V-notch for various opening angles $2\alpha=45^{\circ},90^{\circ},135^{\circ}$. As can be noted, the current density close to the notch tip is singular of order $1/q-1$ with $q=\left(2\pi-2\alpha\right)/\pi$; (b) Cartesian components of the current density as a function of the curvilinear coordinate, $u$, for $v=3$.}
	\label{f12}
\end{figure*}

The current density can be calculated taking advantage of equation (\ref{eq2.13}):
\begin{equation}
J_x-\mbox{i}J_y=\psi\left[\frac{\mbox{d}z\left(\xi\right)}{\mbox{d}z\xi}\right]^{-1}=\psi\frac{\left(\xi^2-1\right)^{1/2-\alpha/\pi}}{\xi^{1-2\alpha/\pi}}
\label{eq2.47bisbis}
\end{equation}
which can be used to calculate the following Cartesian components of the current density in curvilinear coordinates:
\begin{equation}
J_x\left(u,v\right)=\psi\frac{\left[\left(u^2-v^2-1\right)^2+4u^2v^2\right]^{1/4-\alpha/2\pi}}{\left(u^2+v^2\right)^{1/2-\alpha/\pi}}\cos\eta
\label{eq2.47bisbisbis}
\end{equation}
\begin{equation}
J_y\left(u,v\right)=-\psi\frac{\left[\left(u^2-v^2-1\right)^2+4u^2v^2\right]^{1/4-\alpha/2\pi}}{\left(u^2+v^2\right)^{1/2-\alpha/\pi}}\sin\eta
\label{eq2.47bisbisbisbis}
\end{equation}
where $\eta=\left(1/2-\alpha/\pi\right)\Arg\left(u^2-v^2-1+2\mbox{i}uv\right)-\left(1-2\alpha/\pi\right)\Arg\left(u+\mbox{i}v\right)$. Figure \ref{f11}b shows the distribution of the magnitude of the current density for $\psi=1$ A/mm$^2$ whereas Figure \ref{f12}a shows a double-logarithmic plot of the Cartesian component of the current density, $J_x$, along the bisector of the notch calculated by equation (\ref{eq2.47bisbisbis}) and FEA. For the numerical analysis, a plate of height $H=1000$ mm, width $D=500$ mm, notch depth $a=10$ mm, and opening angles $2\alpha=45^{\circ}$, $90^{\circ}$, and $135^{\circ}$ was considered. As can be noted, the current density close to the notch tip is singular of order $1/q-1$ with $q=\left(2\pi-2\alpha\right)/\pi$. The theoretical predictions (solid lines) and numerical results (symbols) are in excellent agreement.

\subsection{Rectangular notch}
\begin{figure*}
	\center
	\includegraphics[trim=0cm 0cm 0cm 0cm, clip=true,clip=true,width = 1\textwidth]{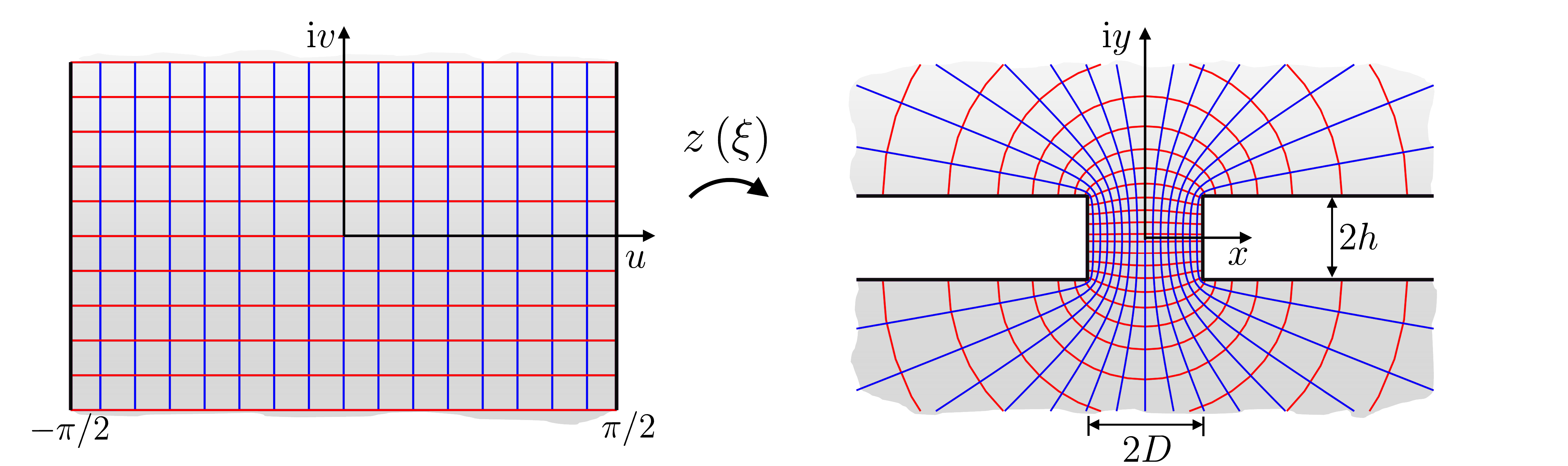}
	\caption{Schematic representation of the conformal mapping in equation (\ref{eq2.48}) which transforms the infinite strip $\left[-\pi/2,\pi/2\right]\times\left(-\infty,\infty\right)$ to an infinite domain featuring two rectangular notches of width $2h$ of infinite depth with a finite ligament $2D$.}
	\label{f13}
\end{figure*}

Let us consider a rectangular cutout in an infinite plate subjected to a remote current density or a potential differential. The conformal mapping to describe the notch profile can be chosen as:
\begin{equation}
z\left(\xi\right)=D E_2\left(\xi,\frac{b}{a}\right)/E_{2c}\left(\frac{b}{a}\right)
\label{eq2.48}
\end{equation}
\\
where $E_2\left(x,y\right)=\int_0^x{\sqrt{1-y^2\sin^2\theta \mbox{d}\theta}}$ and $E_{2c}\left(y\right)=E_2\left(\pi/2,y\right)$ are the incomplete and complete elliptic integrals of the second kind respectively \cite{AbrSte72} and $D$ is a parameter controlling the length of the net section of the plate (Figure \ref{f13}). The notches are described by the condition $u=\pm \pi/2$ while the width of the notch, $h$, can be expressed as:
\begin{equation}
h=\operatorname{Im} \left\{\lim_{v\rightarrow \infty} DE_2\left(\pi/2+iv,b/a\right)/E_{2c}\left(b/a\right) \right\}
\label{eq2.49}
\end{equation}
upon the condition that $b\le a$. Then, the ratio between the width of the notch and the ligament can be calculated explicitly by the following equation \cite{SalZap16}:

\begin{equation}
	\frac{h}{D}=-2\sqrt{\frac{b}{a}}\frac{\operatorname{Re}\left\{E_{2c}\left(1-a/b\right)-a/bE_{1c}\left(1-a/b\right)\right\}-\operatorname{Im}\left\{E_{2c}\left(a/b\right)-\left(1-a/b\right)E_{1c}\left(a/b\right)\right\}}{E_{2c}\left(b/a\right)}
	\label{eq2.50}
\end{equation}

where $E_{1c}\left(y\right)=\int_0^{\pi/2}{\frac{\mbox{d}\theta}{\sqrt{1-y^2\sin^2\theta}}}$ is the complete integral of the first kind \cite{AbrSte72}.
\subsubsection{Optimal fiber paths}
As shown in Figure \ref{f13}, the boundary of the rectangular cutout is described by the condition $u=\pm \pi/2$. According to Section \ref{remarks}, the optimal fiber paths can be found by means of equation (\ref{eq2.13}) imposing $x\left(s,u_c,v_c\right)=\operatorname{Re}\left[\left(z \circ\xi\right)\left(s,u_c,v_c\right)\right]$ and $y\left(s,u_c,v_c\right)=\operatorname{Im}\left[\left(z \circ\xi\right)\left(s,u_c,v_c\right)\right]$. The dashed lines in Figure \ref{f14}a show a schematic representation of the fiber paths that minimize the electrical resistance of the body, coinciding with the field lines.
\subsubsection{Electric potential}
The electric potential can be calculated by means of equation (\ref{eq2.9}) although, in this case, the solution in Cartesian coordinates cannot be found easily in closed-form. The implicit description of the potential reads:
\begin{equation}
\varphi\left(x,y\right)=\zeta\operatorname{Im}\left[\xi\left(z\right)\right]
\label{eq2.51}
\end{equation}
The representation of equation (\ref{eq2.51}) in curvilinear coordinates is, of course, trivial. Figure \ref{f14}b show the contour plot of the electric potential for $\zeta=1$ V.   

\subsubsection{Current density}
\begin{figure*}
	\center
	\includegraphics[trim=0cm 0cm 0cm 0cm, clip=true,clip=true,width = 1\textwidth]{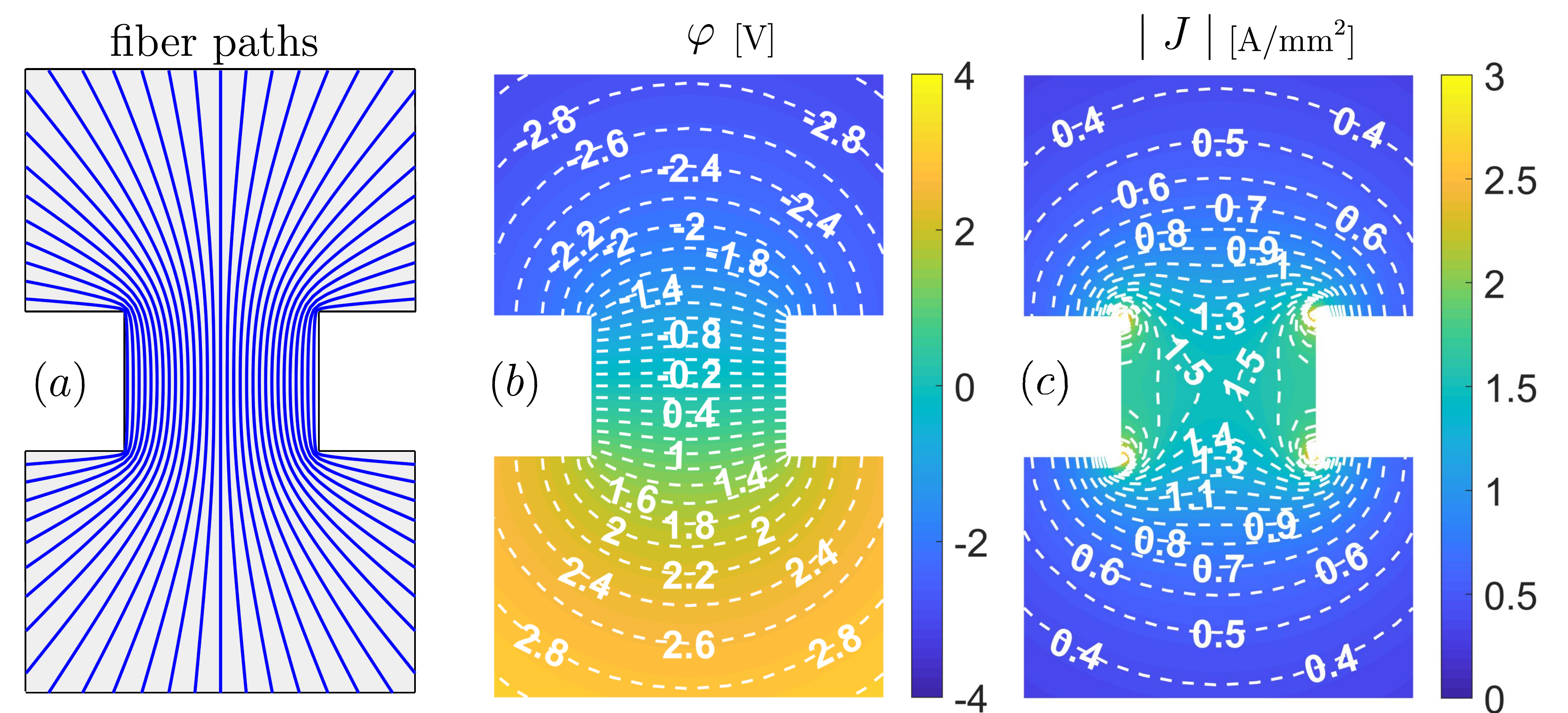}
	\caption{Electrostatic solution for an infinite body featuring curvilinear transverse isotropy weakened by a deep rectangular cutout: (a) fiber path of minimum electrical resistance; (b) electric potential field; (c) magnitude of the electric current density. The notch is described by equation (\ref{eq2.48}) leading to a notch width of $2h=1.45$ mm and a ligament length of $2D=2$ mm.}
	\label{f14}
\end{figure*}

The current density can be calculated from the first derivative of $\xi=\xi\left(z\right)$. However, since this function is difficult to calculate explicitly in this case, one can leverage equation (\ref{eq2.11}) and the derivative of equation (\ref{eq2.48}), obtaining the following equation:
\begin{equation}
J_y+\mbox{i}J_x=\psi\left[\frac{\mbox{d}z\left(\xi\right)}{\mbox{d}\xi}\right]^{-1}=\frac{\psi}{D}\frac{E_{2c}\left(b/a\right)}{\sqrt{1-b/a\sin^2\left(\xi\right)}}
\label{eq2.52}
\end{equation}
which can be used to calculate the Cartesian components of the current density in curvilinear coordinates:

\begin{equation}
	J_y\left(u,v\right)=\psi\frac{\operatorname{Im}\left[E_{2c}\left(\frac{b}{a}\right)\right] \sin \left(\frac{1}{2} \eta\right)+\operatorname{Re}\left[E_{2c}\left(\frac{b}{a}\right)\right] \cos \left(\frac{1}{2} \eta\right)}{\sqrt[4]{\frac{4 b^2 \sin ^2(u) \cos ^2(u) \sinh ^2(v) \cosh ^2(v)}{a^2}+\left[1-b\frac{ \sin ^2(u) \cosh ^2(v)-\cos ^2(u) \sinh ^2(v)}{a}\right]^2}}
	\label{eq2.53}
\end{equation}
\begin{equation}
	J_y\left(u,v\right)=\psi\frac{\operatorname{Im}\left[E_{2c}\left(\frac{b}{a}\right)\right] \cos \left(\frac{1}{2} \eta\right)-\operatorname{Re}\left[E_{2c}\left(\frac{b}{a}\right)\right] \sin \left(\frac{1}{2} \eta\right)}{\sqrt[4]{\frac{4 b^2 \sin ^2(u) \cos ^2(u) \sinh ^2(v) \cosh ^2(v)}{a^2}+\left[1-b\frac{ \sin ^2(u) \cosh ^2(v)-\cos ^2(u) \sinh ^2(v)}{a}\right]^2}}
	\label{eq2.54}
\end{equation}

where $\eta=\Arg \left[1-b/a \sin ^2(u+\mbox{i} v)\right]$. Figure \ref{f14}c shows the distribution of the magnitude of the current density for $\psi=1$ A/mm$^2$ whereas Figure \ref{f12}b shows a comparison between the Cartesian components of the current density along the curve $v=3$ calculated by equations (\ref{eq2.53}) and (\ref{eq2.54}) and FEA. For the numerical analysis, a plate of height $H=1000$ mm, width $W=500$ mm, ligament length $2D=150$ mm was considered. Again, it is worth noting that the theoretical and numerical results are in excellent agreement.

\subsection{Electrical resistance of CTI systems vs. traditional composites}
\begin{figure*}
	\center
	\includegraphics[trim=0cm 0cm 0cm 0cm, clip=true,clip=true,width = 0.9\textwidth]{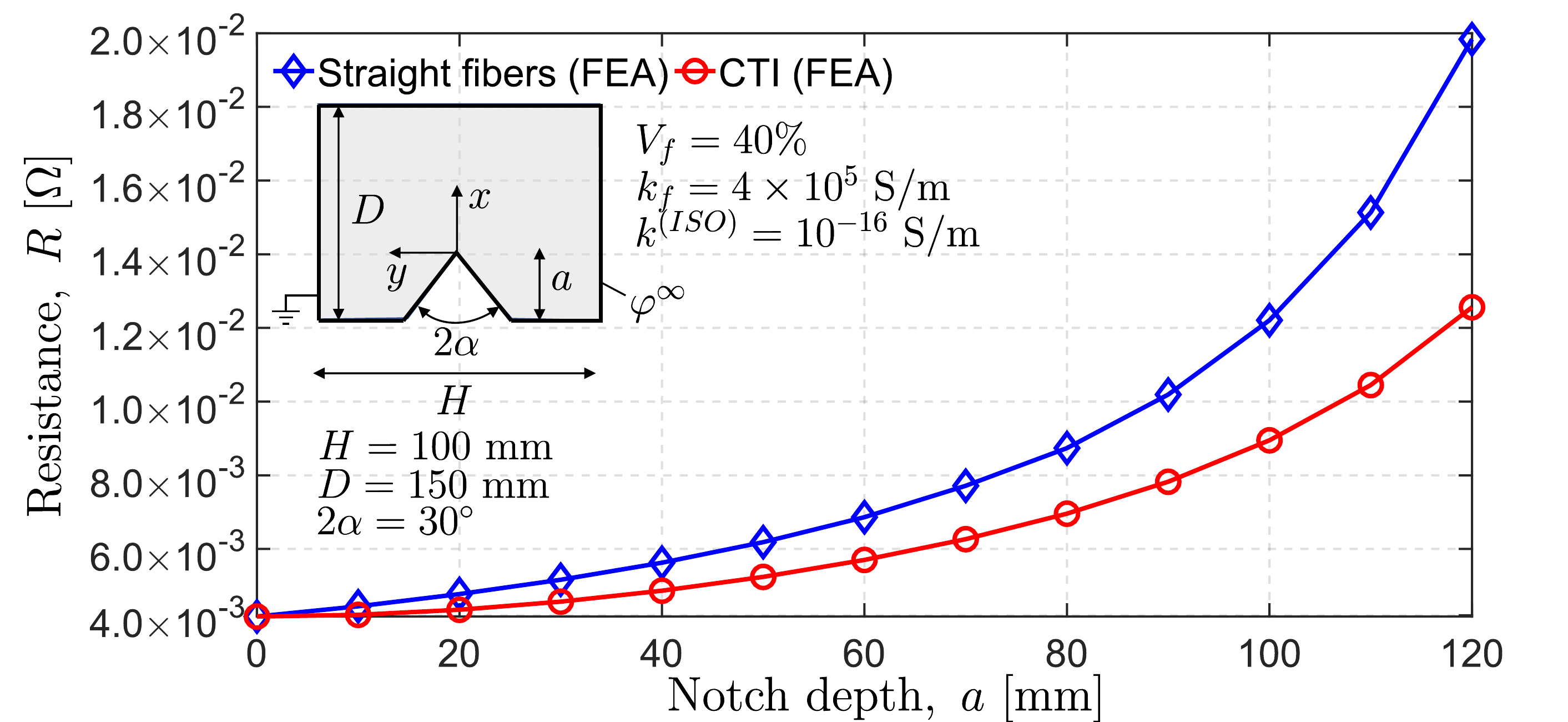}
	\caption{Electrical resistance as a function of notch depth, $a$, for a finite plate featuring a sharp V-notch: straight fibers vs curvilinear fibers. All the material parameters are kept constant. The use of curvilinear fibers leads to lower electrical resistances for a given depth. The advantage of CTI is evident when the notch is deep and the electric field is significantly perturbed.}
	\label{f15}
\end{figure*}

In the foregoing sections, a theoretical framework for the closed-form solution of linear electrostatic problems in media featuring \textit{Curvilinear Transverse Isotropy} (CTI) was proposed. The analytical formulation enabled the derivation of equation (\ref{eq2.36}) which allows the explicit calculation of the electrical resistance of the CTI system for any fiber orientation. Such an expression can be simplified to equation (\ref{eq2.37}) when the fibers follow the field lines, which was shown to be the case optimizing the electrical conductivity. Leveraging this result, it is interesting to compare the electrical resistance of a plate notched by a finite sharp V-notch of depth $a$, in case the medium exhibits CTI or the fibers follow a straight longitudinal path. The latter case is representative of the electrical conductivity of a traditional fiber composite plate reinforced by carbon fibers. The comparison is shown in Figure \ref{f15} which provides the electrical resistance as a function of the notch depth for a constant opening angle $2\alpha=30^{\circ}$. The analyzed plate has a height $H=100$ mm, a width $D=150$ mm whereas a unit thickness was considered. A fiber volume fraction $V_f=40\%$, fiber electrical conductivity $k_f=4\times 10^5$ S/m and matrix electrical conductivity $k^{(ISO)}=10^{-16}$ S/m were used in equations (\ref{eq2.38}a,b) to calculate the electric properties of the medium in its material coordinate system for both configurations. The electrical resistance was calculated leveraging FEA imposing a unitary potential difference at the left and right boundaries of the plate while the rest of the boundary was considered isolated ($\textbf{J}\cdot \textbf{n}=J_n=0$). This plot clearly highlights the superior performance of CTI systems compared to composites featuring straight fibers in terms of electrical resistance. In fact, for a given notch depth $a$, $R$ is always lower for CTI than for traditional composites, the difference growing with increasing notch depth and reaching almost $60\%$ for $a=0.8D$. This is because, for deeper notches, the perturbation of the electric field from the un-notched solution becomes larger and larger and thus the morphology of the fibers starts to play a key role.

\subsection{Comments on the applicability of the theoretical framework}
The theoretical framework described in the foregoing sections and the related corollaries enabling the straightforward calculation of the electrical resistance of CTI plates depend on the fulfillment of specific hypotheses. The first is that separation of length-scales can be assumed so that the material can be treated as homogeneous and transversely isotropic along the curvilinear paths. It is also assumed that Dirichlet or von Neumann boundary conditions are ascribed to portions of the boundary where one of the curvilinear coordinates describing the domain is constant. \textit{This latter requirement seems particularly restricting, yet, it is not}. In fact, the literature abounds of handbooks and collections of conformal mappings satisfying such conditions and enabling the description of a plethora of domains \cite{Math94,Iva94,Kythe18}[e.g]. In case of polygonal domains, the map can be even constructed geometrically \cite{DriTre02}. Finally, it should be noted that equation (\ref{eq2.37}) can be used even if the conformal mapping is not known or it is too complicated to allow a closed-form solution. In fact, it is worth recalling that the electric potential is an harmonic function and it is constant where Dirichlet boundary conditions are applied. On the other hand, insulating boundaries are always described by field lines of constant electric field since along $\partial\Omega_{n,0}$ the normal derivative of the electric potential is null: $\nabla\varphi\cdot \textbf{n}=0$. Accordingly, the electric potential $\varphi$ and the the related orthogonal field lines represent the real and imaginary part of the conformal mapping satisfying exactly the hypotheses. Of course, these quantities are also the unknowns of the problem. Hence, their use does not provide much information if the closed-form solution of the electrostatic potential and the related electric field is of interest. However, the fact that these functions exist, confirms the general validity of equation (\ref{eq2.37}). In other words, this equation can always be used to correlate the electric resistance of CTI plates to the isotropic ones.
Finally, it worth noting that the hypotheses on the description of the Dirichlet and von Neumann boundary conditions could be further relaxed. In such a case, however, the solution of the governing equation could be obtained only by means of a Laurent series expansion, not a closed-form solution.

\section{Conclusion}
This work proposed a general framework for the exact, closed-form solution of linear electrostatic problems in media featuring \textit{Curvilinear Transverse Isotropy} (CTI), i.e. materials in which a matrix phase is reinforced by fibers following desired curvilinear paths. The approach enables the exact calculation of the electric potential, the electric field, the electrical resistance as well as the family of fiber paths maximizing the electrical conductivity. Based on the results obtained in this study, the following conclusions can be elaborated:

\begin{enumerate}%[(a)]
	
	\item to maximize the electrical conductivity of the system, the fiber paths should follow the field lines of the related isotropic electrostatic problem;

	\item in case the curvilinear fibers are rotated of a constant angle $\rho_0$ with respect to the field lines, the current density of the CTI system coincides with the isotropic case but for a multiplying constant that depends on the boundary conditions of the problem and the properties of the constituents. In such a case, the curvilinear components of the current density can be described by a complex analytic function which was proven to be the first derivative of the conformal map required to describe the domain of the problem, $\Omega\cup\partial\Omega$;

	\item thanks to the foregoing result, the solution of the electrostatic problem comes down to simply identifying the conformal mapping properly describing the domain;

	\item in general, while the electric current densities are formally similar, the electric potential and electric field in CTI and isotropic materials take different expressions. However, when the fibers follow the paths maximizing the electrical conductivity ($\rho_0=0$), all the quantities of interest take the same form;

	\item the theoretical framework enabled the formulation of a strikingly simple equation which relates the electrical resistance of the isotropic case to the one of the CTI system. This equation depends only on the properties of the constituents, the volume fraction of fibers and fiber angle;

	\item the foregoing result is particularly important for the design of CTI components. First, it allows an easy estimate of the reduction of electrical resistance introduced by the addition of conductive fibers. Since the expression is derived in closed-form, it is possible to know, explicitly, the volume fraction of fibers and fiber conductivity required to achieve a desired electrical resistance. Second, the equation simplifies significantly the numerical simulation of CTI structures. In fact, one can simply simulate the isotropic case using standard FE solvers and then use the proposed equation to infer the performance of the related CTI system avoiding the need for advanced user subroutines to describe the fibers and the inaccuracy introduced by the generally poor description of the fiber paths in commercial FE solvers;

	\item it was shown that CTI systems guarantee a superior performance compared to e.g. structures featuring straight fibers, for the same properties of the constituents. In the presence of a notch, the advantage of CTIs over traditional systems becomes more and more significant with increasing notch depths or, in other words, increasing perturbation of the current density field compared to the un-notched case;

	\item all the foregoing results are of utmost importance for the design of material systems leveraging \textit{Curvilinear Transverse Isotropy} (CTI) to achieve superior electrical or thermal conductivity. Possible applications include the use of this technology to enable damage self-sensing based on resistivity measurements in typically insulating materials such as e.g. polymers. In such a case, the addition of conductive, curvilinear fibers can lead to sufficient levels of conductivity. Further, the fiber paths can be designed to e.g. maximize the change in electrical resistance in the presence of a crack initiating from a notch. Ongoing work by the authors towards this direction is providing outstanding results. 
\end{enumerate}

\section*{Acknowledgement}
%\begin{acks}
	Marco Salviato acknowledges the financial support from the Haythornthwaite Foundation
	through the \textit{ASME Haythornthwaite Young Investigator Award}. This work was also partially supported by the Joint Center for Aerospace Technology Innovation (JCATI) of the state of Washington, USA, through the project titled \textit{``Manufacturing and Buckling Study of Curved Steered Sandwich Panels using Automated Fiber Placement (AFP) and Out-Of-Autoclave (OOA) for Space Launch Vehicles"}.
%\end{acks}

%\begin{thebibliography}{99}

\bibliographystyle{vancouver}
\bibliography{CTI}

%\end{thebibliography}
\end{document}